\documentclass[a4paper,12pt]{article}
\pdfoutput=1
\usepackage{amssymb}
\usepackage{amsmath}
\usepackage{epsfig}
\usepackage{graphicx}
\usepackage{slashed}
\usepackage{color}
\usepackage{ulem}

\makeatletter
\renewcommand{\theequation}{\thesection.\arabic{equation}}
\@addtoreset{equation}{section}
\makeatother
\def\be{\begin{equation}}
\def\ee{\end{equation}}
\def\bea{\begin{eqnarray}}
\def\eea{\end{eqnarray}}
\def\eq{\begin{eqnarray}}
\def\eqx{\end{eqnarray}}
\def\nn{\nonumber \\}
\def\({\left(}
\def\){\right)}
\def\<{\left<}
\def\>{\right>}

\def\be{\begin{equation}}
\def\ee{\end{equation}}
\def\ben{\begin{eqnarray}}
\def\een{\end{eqnarray}}
\def\({\left(}
\def\){\right)}
\def\<{\left<}
\def\>{\right>}
\def\!{\right|}
\def\|{\left|}

\def\[{\left[}
\def\]{\right]}

\def\+{\bar}

\usepackage{hyperref}
\hypersetup{colorlinks,%
  citecolor=blue,%
  linkcolor=blue,%
  pdftex}

\begin{document}

\begin{titlepage}
\vskip1cm
\begin{flushright}
\end{flushright}
\vskip0.25cm
\centerline{
\bf \large 
Bulk 
 View of %
Teleportation and Traversable 
Wormholes  
} 
\vskip1cm \centerline{ \textsc{
 Dongsu Bak,$^{\, \tt a}$  Chanju Kim,$^{\, \tt b}$ Sang-Heon Yi$^{\, \tt a}$} }
\vspace{1cm} 
\centerline{\sl  a) Physics Department,
University of Seoul, Seoul 02504 \rm KOREA}
 \vskip0.3cm
 \centerline{\sl b) Department of Physics, Ewha Womans University,
  Seoul 03760 \rm KOREA}
 \vskip0.4cm
 \centerline{
\tt{(dsbak@uos.ac.kr,\,cjkim@ewha.ac.kr,\,shyi704@uos.ac.kr})} 
  \vspace{2cm}
\centerline{ABSTRACT} \vspace{0.75cm} 
{
\noindent We construct detailed AdS$_2$ gravity solutions describing the 
teleportation through a traversable wormhole
sending a state from one side of the wormhole to the other.
The traversable wormhole is realized by turning on a double trace interaction
that couples the two boundaries of an eternal AdS$_2$ black hole. 
The horizon radius or the entropy of the black hole is reduced 
consistently with the boundary computation of the energy change,
confirming the black hole first law.
To describe teleportee states traveling through the wormhole,
we construct Janus deformations which make the Hamiltonians
of left-right boundaries differ from each other by turning on 
exact marginal operators. Combining explicitly the traversable wormhole 
solution and the teleportee states, we present a complete bulk picture 
of the teleportation in the context of ER=EPR. The traversability
of the wormhole is not lost to the leading order of the deformation parameter.
We also consider solutions where the teleportee meets the matter thrown
from the other side during teleportation, in accordance with the assertion
that the bulk wormhole is experimentally observable.
}

\end{titlepage}

%
\section{Introduction}
There are some renewed interests in AdS$_{2}$ space, inspired by the proposal for its correspondence with the four-Fermi  random interaction model, known as the SYK model~\cite{SYK} {(See~\cite{Sarosi:2017ykf} for a review)}. Historically, AdS$_{2}$ space has received the attention as the essential part in the near horizon of the extremal black holes. Since the temperature of extremal black holes vanishes and  those black holes do not emit the Hawking radiation, those are regarded as stable objects with mass gap, providing an ideal test ground for various methodologies of the microscopic counting of black hole entropy. One might anticipate the concrete realization of ideas  or analytic computations on black holes  in the context of AdS$_{2}$/CFT$_{1}$ correspondence. On the contrary, it turns out to be a bit twisted to construct a meaningful gravity  theory on two-dimensional spacetime, since pure Einstein theory becomes topological on two dimensions. Recent developments in the correspondence utilize the freedom in the boundary degrees in the two-dimensional gravity, and so the nearly-AdS$_{2}$ space is taken as the bulk background. 

Another interesting aspect of AdS$_{2}$ space is that it has two boundaries different from the single boundary in its higher dimensional cousins, which may put a  hurdle on the direct adaptation of methods in higher dimensional case. However,  even in higher dimensional AdS case, it has been known that the eternal AdS black holes provide two boundaries and can naturally be identified with the highly-entangled, so-called,  thermo field double states (TFD) in the finite temperature field theory~\cite{Israel:1967wq,Maldacena:2001kr}. Recently, this aspect of the existence of two boundaries  in eternal AdS black holes  and its correspondence with TFD has led to an interesting bulk realization of quantum teleportation: traversable wormhole~\cite{Susskind:2014yaa,Susskind:2016jjb,Gao:2016bin,Maldacena:2017axo,Susskind:2017nto,Maldacena:2018lmt}.  By turning on the double trace interaction between two boundaries with a negative energy, it is explicitly shown that the average null-energy condition in the bulk is violated and so wormholes could be traversable. This bulk geometry is argued to be interpreted as the gravity realization of the quantum teleportation in the dual theory. Though the turned-on interaction between two boundaries is taken to be very small admitting its perturbative treatment, it is argued that the bulk deformation caused by the back-reaction  renders wormholes as traversable ones.

In this paper, we consider the two-dimensional Einstein-dilaton model with a scalar field and investigate the concrete bulk dilaton dynamics. In this model, one can show by the explicit computation that the dilaton dynamics by the boundary interaction cause the position of the singularity of black holes is moved in a way that the wormhole becomes traversable.  One can also show that the horizon radius or the entropy of black holes is reduced consistently with the black hole 1st law. Furthermore, we consider the thermalization and Janus deformation of black holes and show that it could be combined with the two boundary interaction consistently. This could be regarded as the complete bulk realization of  the quantum teleportation. 

Although this bulk description can be made fully consistent in its own right, there is in general an extra back-reaction effect in identification of its corresponding 
boundary system. Once there are any excitations from an AdS$_2$ black hole, identification of the left-right (L-R) boundary time coordinates $t_{L/R}$ as 
a function of the bulk time coordinate $t$ at  regulated
 L-R boundaries becomes nontrivial. Without any excitations above the thermal vacuum, one has
simply $t(t_{L/R})= t_{L/R}$.  On the other hand, if the system is excited, 
this (reparameterization) dynamics becomes nontrivial as was emphasized in Ref.~\cite{Maldacena:2016upp}.
In this note, we shall show the consistency of our bulk description with that of the boundary side only to the leading order. 
Of course the full identification of the correspondence requires the formulation introduced in Ref.~\cite{Maldacena:2016upp}, which we shall not attempt to do in this note.  See also Refs.~\cite{Maldacena:2017axo,Maldacena:2018lmt} for the account of teleportation in this direction.

This paper is organized as follows. In Section \ref{sec2}, we present our model and summarize basic black hole solutions and their basic properties. In Section \ref{sec3}, we consider the scalar field perturbation of black holes and its thermalization. In Section  \ref{sec4}, we provide a specific time-dependent Janus deformation of AdS$_{2}$ black holes and show that one cannot send signal from one boundary to the other in this case. In Sections  \ref{sec5} and  \ref{sec6}, we consider the double trace deformation between two boundaries and  show that it renders the wormhole to be traversable with the explicit entropy/temperature reduction. In Section  \ref{sec7}, we combine our results in previous sections and provide the complete bulk picture dual to the quantum teleportation. We conclude in Section  \ref{sec8} with some discussion. Various formulae are relegated to Appendices.

\section{Two-dimensional dilaton gravity 
}\label{sec2}
We begin with the 2d dilaton gravity  in Euclidean space~\cite{Jackiw:1984je,Teitelboim:1983ux,Almheiri:2014cka}
\bea
I=I_{top}-{1\over 16\pi G}\int_M d^2 x \sqrt{g}\, \phi \left( R+\frac{2}{\ell^2}\right) +I_M(g, \chi)\,,
\label{euclidaction}
\eea
where 
\begin{align}    \label{}
 I_{top}&=  -{\phi_0\over 16\pi G}\int_M d^2 x \sqrt{g}  R\,,  \nonumber \\
I_M &= \frac{1}{2}\int_M d^2 x \sqrt{g} \left( \nabla \chi \cdot \nabla \chi + m^2 \chi^2 \right)\,.
\end{align}
Below we shall evaluate the above action on shell, which  would diverge if the boundary is taken  at  infinity. 
For its regularization, we introduce  a cutoff surface $\partial M$ near infinity. This requires adding  surface terms
\begin{equation} \label{}
I_{surf} =   -{1\over 8\pi G}\int_{\partial M} \sqrt{\gamma}\, (\phi_0 + \phi ) \, K \,,
\end{equation}
where $\gamma_{ij}$ and $K$  
denote the induced metric and  the extrinsic curvature. 
Then the renormalized  action  (obtained by adding the counter terms)
 corresponds to the free energy multiplied by $\beta$,
\bea
I_{ren} =\beta F= -\log Z \,,
\eea
where $Z$ is the partition function of the dual  quantum mechanical system.

The corresponding Lorentzian action takes the form
\bea
I=I_{top}+{1\over 16\pi G}\int_M d^2 x \sqrt{-g}\, \phi \left( R+\frac{2}{\ell^2}\right) +I_M(g, \chi)\,,
\eea
where 
\begin{align}    \label{}
I_{top}&=  {\phi_0\over 16\pi G}\int_M d^2 x \sqrt{-g} \,  R\,,  \nonumber \\
I_M &= -\frac{1}{2}\int_M d^2 x \sqrt{-g} \left( \nabla \chi \cdot \nabla \chi + m^2 \chi^2 \right)\,.
\end{align}
The equations of motion read
\begin{align}    \label{phieq}
&R+\frac{2}{\ell^2}=0\,, \nonumber \\
& \nabla^2 \chi -m^2 \chi =0\,, \nonumber \\
& \nabla_a \nabla_b \phi -g_{ab} \nabla^2 \phi + g_{ab} \phi = - 8 \pi G T_{ab}\,,
\end{align}
where
\begin{equation} \label{}
T_{ab} = \nabla_a \chi \nabla_b \chi  -\frac{1}{2} g_{ab} \left( \nabla \chi \cdot \nabla \chi + m^2 \chi^2 \right)\,.
\end{equation}
%

\begin{figure}
\vskip-1cm
\begin{center}
\includegraphics[width=6.3cm,clip]{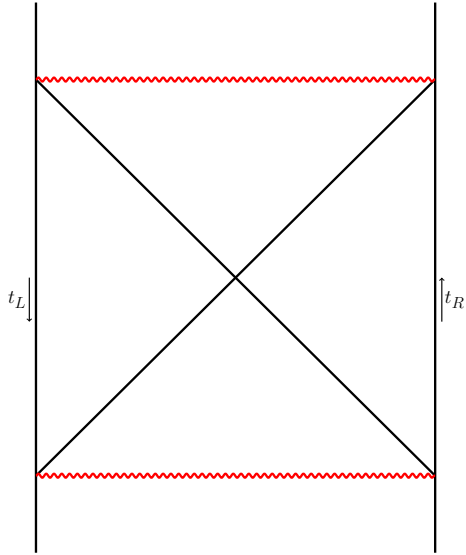}
\end{center}
\vskip-1cm
\caption{
\label{fig02} We draw the Penrose diagram for the AdS$_2$ black hole with $b=1$ in $(\tau,\mu)$ space where the wiggly red lines represent the location 
of singularity.
}
\end{figure}

Any AdS$_2$ space can be realized by the global AdS space whose metric is given by
\begin{equation} \label{}
ds^2 =\frac{\ell^2}{\cos^2 \mu} \left(-d\tau^2 + d\mu^2  \right)\,,
\end{equation}
where $\mu$ is ranged over $[-\frac{\pi}{2},\frac{\pi}{2}]$.
The most general vacuum solution for the dilaton field is given by
\begin{equation} 
\phi= \frac{1}{\cos \mu} \left( \alpha_0 \cos \tau + \alpha_1 \sin \tau \right) + \alpha_2   \frac{\sin \mu}{\cos \mu}\,.
\label{hom}
\end{equation}
%
Using the translational isometry along $\tau$ direction, one may set $\alpha_1$ to zero without loss of generality. 
We shall parameterize the dilaton field by
\bea \label{DilatonSol0}
\phi= \phi_{BH}(L,b,\tau_{B})\equiv \bar\phi \, L\,\,\frac{(b+b^{-1}) \cos (\tau-\tau_B) -(b-b^{-1}) \sin \mu}{2 \cos \mu} \,,
\label{dilaton}
\eea 
where we choose $b \ge 0$.
By
the coordinate transformation 
\begin{align}    \label{}
\frac{r}{L} &= \frac{(b+b^{-1}) \cos (\tau-\tau_B) -(b-b^{-1}) \sin \mu}{2\cos \mu}\,,  \nonumber \\
 \tanh \frac{t L }{\ell^2} &=\frac{2\sin (\tau-\tau_B)}{(b+b^{-1}) \sin \mu -(b-b^{-1}) \cos (\tau-\tau_B)}\,,
 \label{coorb}
\end{align}
one is led to the corresponding AdS black hole metric
\begin{equation} \label{btz}
ds^2= - \frac{r^2-L^2}{\ell^2} dt^2+ \frac{\ell^2}{r^2-L^2} dr^2\,,
\end{equation}
with
\begin{equation} \label{}
\phi= \bar\phi \, r\,.
\end{equation}
The  Penrose diagram for the above black hole with $b=1$ is depicted  in Figure \ref{fig02}. 
They in general describe two-sided AdS black holes. The location of singularity is defined by the curve $\Phi^2\equiv \phi_0 +\phi =0$ in 
the above dilaton gravity  where  $\Phi^2$ might be viewed as characterizing the size of  ``transverse space"~\cite{Almheiri:2014cka}.  In the figure, 
we set $\phi_0 =0$ for definiteness.
%

We now compute the free energy. For this, we go to the Euclidean space where the AdS solution read 
\begin{equation} ds^2=  \frac{r^2-L^2}{\ell^2} dt_E^2+ \frac{\ell^2}{r^2-L^2} dr^2 
\label{btz1}
\end{equation}
with
\begin{equation} \label{}
\phi= \bar\phi \, r\,.
\end{equation}
%
The Gibbons-Hawking temperature can be identified as
\begin{equation} \label{}
T= \frac{1}{2\pi} \frac{L}{\ell^2}\,.
\end{equation}
Note that the Euler number is defined by
\begin{equation} \label{}
\chi = \frac{1}{4\pi} \left[\, \int_M d^2 x \sqrt{g}  R + 2 \int_{\partial M} dx \sqrt{\gamma} K \,
\right]\,,
\end{equation}
which does not require any counter term.
The (renormalized) topological term can be evaluated as
\begin{equation} \label{}
I^{top}_{ren} = -\frac{\phi_0}{4G}\,,
\end{equation}
where we used the fact that $\chi=1$ for the thermal disk geometry. For the evaluation of the rest terms, we cutoff the bulk at
\begin{equation} \label{}
\frac{r}{L}
=\frac{1}{\delta}\,.
\end{equation}
%
Note that  the second term in (\ref{euclidaction}) is zero on-shell. Then the remaining term becomes
\begin{equation} \label{}
\Delta I_{reg} =  -{1\over 8\pi G}\int_{\partial M} \sqrt{\gamma}\, \phi \, K = -\frac{\bar\phi}{8\pi G} 
\beta \frac{L^2}{\ell^2} \frac{1}{\delta^2}\,.
\end{equation}
For the renormalization one has to go to the Fefferman-Graham coordinates. The metric in (\ref{btz1}) becomes
\begin{equation} \label{}
ds^2 =\frac{\ell^2}{z^2} dz^2 + \frac{r^2(z)-L^2}{\ell^2} dt_E^2\,, 
\end{equation}
where
\begin{equation} \label{}
\frac{r}{L}
  =\frac{1+z^2}{2z}\,.
\end{equation}
Therefore the cutoff $\epsilon$ in $z$ coordinate is related to $\delta$ by
\begin{equation} \label{}
\delta = \frac{2\epsilon}{1+\epsilon^2}\,,
\end{equation}
and then
\begin{equation} \label{}
\Delta I_{reg} = -\frac{\bar\phi}{8\pi G} 
\beta \frac{L^2}{\ell^2} \frac{(1+\epsilon^4 + 2\epsilon^2)}{4 \epsilon^2}\,.
\end{equation}
Subtracting the divergent term from $\Delta I_{reg}$ by a counter term $\sim \int_{z=\epsilon}\sqrt{\gamma} \, \phi $, one has
\begin{equation} \label{}
\Delta I_{ren} =-\frac{\bar\phi}{16\pi G} 
\beta \frac{L^2}{\ell^2}  = -\frac{{\cal C}}{2} T\,,
\end{equation}
where
\begin{equation} \label{}
{\cal C} =\frac{\pi \bar\phi \ell^2}{2 G}\,.
\end{equation}
Thus the free energy becomes
\begin{equation} \label{}
F= - S_0 T  -\frac{{\cal C}}{2} T^2\,,
\end{equation}
with
\bea
S_0 =\frac{\phi_0}{4G}\,.
\eea
The entropy  and energy are then
\begin{align}    
S&= S_0 +{\cal C} T\,,   \\  
E &= \frac{1}{2} {\cal C}T^2\,.
\label{energybh} 
\end{align}
We note that the deformation in $b$ does not play any role in the thermodynamics. In addition, note that 
the entropy can be written as a Beckenstein formula
\begin{equation} \label{}
S= \frac{\phi_0 +\bar\phi\, L} {4G}\,.
\end{equation}
As alluded earlier, 
we shall ignore the effect of $\phi_0$ by setting it to zero since we are not interested in this part of the black hole physics
in the following.  
Below we shall focus on the  $b=1$ case as our initial unperturbed system in constructing wormhole solutions. Of course, this  can be relaxed 
to a general value of $b$. 

 The above two-sided black hole in the AdS$_2$ spacetime is dual to the so-called thermofield double~\cite{Takahasi:1974zn} of CFT$_1$, which can be generalized 
to higher dimensions~\cite{Maldacena:2001kr}.
Without deformation  of L-R coupling, the left and the right systems
of $CFT_L \otimes CFT_R$ are decoupled from each other with Hamiltonians $H_L = H_l \otimes 1$ and  $H_R = 1 \otimes H_r$ 
and the corresponding two time 
parameters $t_L $ and $t_R$, respectively. For the thermofield double of CFT, $H_{l}=H_r= H$ where $H$ is the Hamiltonian of a CFT.
The left boundary time $t_L$ runs downward while $t_R$ runs upward  in the AdS space. This identification is consistent with
the coordinate system (\ref{coorb})  
since  the orientation of time direction of $t$ in the left side is reversed  from that of the right side.  Since the orientation of $t_{L}$ is reverse to that of $t_{R}$,  the time evolution of the
full system is given by the operator $e^{i H_L t_L -i H_R t_R}$.   
When we view the full system embedded in a spacetime with extra dimensions with single time evolution, we 
may choose $t_R = -t_L =t$ with (undeformed) Hamiltonian
\bea
H_{tfd}= H_L + H_R = H \otimes 1  + 1\otimes H\,.
\eea
This single time evolution is also relevant when the left and the right systems are coupled, which is  indeed the case with our teleportation  protocol described below.
 The initial unperturbed thermal vacuum state is given by a particularly prepared entangled state \cite{Maldacena:2001kr} 
\begin{equation}  
|\Psi(0) \rangle =\frac{1}{\sqrt{Z}} \sum_{n,n'} \, \langle n | {\cal U} | n' \rangle \, |n' \rangle \otimes | n  \rangle =\frac{1}{\sqrt{Z}} 
\sum_{n} \,  e^{-\frac{\beta}{2} E_n }\, |n\rangle \otimes | n \rangle\,,
\label{initial}
\end{equation}
with a Euclidean evolution operator ${\cal U}={\cal U}_0 \equiv e^{-\frac{\beta}{2}  H}$ and $Z$ denoting the normalization factor. 
The left-right entanglement here is maximal for a given temperature 
$T$.

\section{Thermalization}\label{sec3}
In this section we shall construct various full back-reacted solutions describing thermalization of initial perturbations
of black holes and investigate their general properties. It will be later on used to understand  the dynamics of teleportee through the bulk.  We will set $\ell=1$ in this and the next section. 

In particular, we would like to solve the equations of motion in  (\ref{phieq}). The matter field $\chi$ with mass $m$ is dual  
to a scalar primary operator $O_\Delta (t)$, where its dimension is  
related to the mass by
\bea
\Delta = \frac{1}{2} \left( 1 + \sqrt{1+4m^2}\right)\,,
\eea 
when $m^2 \ge 0$. When $ 0 > m^2  > -1/4 $,   both possibilities of operator dimensions,  
\bea
\Delta=\Delta_\pm= \frac{1}{2} \left( 1 \pm \sqrt{1+4m^2}\right)\,,
\label{deltapm}
\eea 
may be realized\footnote{For the double trace deformation discussed  in 
Section \ref{sec5}, in particular, we shall consider the operator of dimension $\Delta=\Delta_- $ which is ranged over
$(0, \frac{1}{2})$.}.

The scalar field equation can be solved by \cite{Spradlin:1999bn}
\bea
\chi= \sum^\infty_{n=0} c_n \, {\cal N}_n \cos^\Delta \mu \, C_n^\Delta (\sin \mu) \, e^{-i(n+\Delta)\tau 
}  + {\rm c.c.}
\eea
where 
\bea
{\cal N}_n =2^{\Delta -1} \Gamma(\Delta)\sqrt{\frac{\Gamma(n+1)}{\pi \Gamma(n+2\Delta)}}\,,
\eea
and $C^\Delta_n(x)$ denotes the Gegenbauer polynomial defined in \cite{grad}. This bulk solution is dual to the deformation of thermofield initial state with a Euclidean 
evolution operator 
\bea
{\cal U}= e^{-\frac{\beta}{4}  H}  e^{-\sum_n c_n O_n^\Delta}  e^{-\frac{\beta}{4}  H}\,,
\eea
where we add a linear combination of operators $O_n^\Delta$ at the mid-point of the Euclidean evolution where $O_n^\Delta$'s denote the primary operator $O_\Delta$ and its
descendants \cite{Bak:2017xla}.   The Hamiltonian of the boundary system is undeformed on the other hand.

\begin{figure}
\vskip-1cm
\begin{center}
\includegraphics[width=6.3cm,clip]{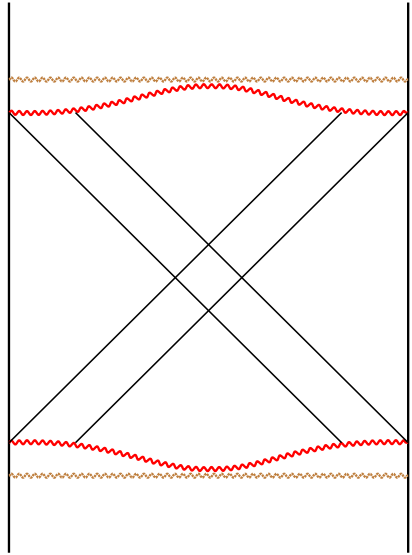}
\end{center}
\vskip-1cm
\caption{
\label{fig021}We depict the deformation of the Penrose diagram with initial perturbation of thermal vacuum.
}
\end{figure}

To see  what this deformation describes, we need to look at the dilaton part whose identification will 
complete the fully back-reacted gravity solution of  (\ref{phieq}).  Here we shall consider only $n=0$ case for the sake of an illustration.
One may write the $n=0$ scalar solution as
\bea
\chi =\gamma  \cos^\Delta \mu \cos \Delta (\tau+ \tau_C )\,,
\eea
without loss of generality. The corresponding dilaton solution can be found as
\begin{align}    \label{}
\phi &=  \bar\phi \, \tilde{L}\,\,\frac{(b+b^{-1})}{2}\frac{\cos \tau }{ \cos \mu}  -\frac{2\pi G\gamma^2 \Delta }{1+2\Delta} \cos^{2\Delta} \mu \cos 2\Delta (\tau+ \tau_C )  \nonumber \\
 &~~~ - 2\pi G\gamma^2 \Delta \cos^{2\Delta} \mu \,\, F \Big(\Delta, 1\,;\,  \frac{1}{2} \,\Big|\, \sin^2 \mu\Big)\,,
\end{align}
where $\tilde{L}=L+ \delta L$, $b=1+\delta b$ and $F(a,b\,;\,c\,|\,z)$ denotes the hypergeometric function \cite{grad}. We have added here a homogeneous solution that is consistent with the symmetry of the perturbation
under 
$\mu \rightarrow -\mu$. To see its asymptotic structure in the region $\mu\rightarrow \pm\frac{\pi}{2}$,  we shall use the following relation
\begin{equation} \label{}
F\Big(\Delta, 1\,;\,  \frac{1}{2}\,\Big|\, \sin^2 \mu\Big) = \frac{1}{1+2\Delta}\,\, F\Big(\Delta, 1\,;\,  \Delta +\frac{3}{2}\,\Big|\, \cos^2 \mu\Big)
 +\frac{\Gamma(\frac{1}{2})\Gamma(\Delta +\frac{1}{2})}{\Gamma(\Delta) \cos^{2\Delta} \mu }|\tan \mu|\,.
\end{equation}
In the asymptotic region,  the solution  becomes 
\bea
\phi = \bar\phi \, \tilde{L}\,\,\frac{(b+b^{-1})}{2}\frac{\cos \tau }{ \cos \mu} 
 - 2\pi G\gamma^2 \frac{\Delta \Gamma(\frac{1}{2})\Gamma(\Delta +\frac{1}{2})}{\Gamma(\Delta) }\frac{|\sin \mu|}{\cos \mu}
  + {\cal O}(\cos^{2\Delta}\mu)\,,
\eea
which will be compared to the black hole solution (\ref{DilatonSol0}). This black hole is symmetric under the exchange of the left and the right as illustrated in 
Figure \ref{fig021}. The new 
temperature of the system
is given by
\bea
\tilde{T}= \frac{1}{2\pi} \frac{\tilde{L}}{\ell^2}\,,
\eea
 and the parameter $b$ on the right side can be determined by
\bea
 \frac{b-b^{-1}}{2}  = \frac{2\pi G\gamma^2}{\bar{\phi}\tilde{L}} \frac{\Delta \Gamma(\frac{1}{2})\Gamma(\Delta +\frac{1}{2})}{\Gamma(\Delta) }\,,
\eea
leading to
\bea
\delta b =\frac{ 2\pi G\gamma^2}{\bar{\phi}L} \frac{\Delta \Gamma(\frac{1}{2})\Gamma(\Delta +\frac{1}{2})}{\Gamma(\Delta) }+{\cal O}(G^2\gamma^4)\,,
\eea
where we assume that $\delta L $ is of order $G\gamma^2$.
Finally the shifts of the singularity in $\tau$ coordinate at the L-R boundaries  can be identified as
\begin{align}    \label{}
\delta \tau_L^\pm = \delta \tau_{R}^\pm &= \mp \arcsin \left[ \frac{ 4\pi G\gamma^2}{\bar{\phi}\tilde{L}(b+b^{-1})} \frac{\Delta \Gamma(\frac{1}{2})\Gamma(\Delta +\frac{1}{2})}{\Gamma(\Delta) } \right] \nonumber \\
& = \mp\frac{ 2\pi G\gamma^2}{\bar{\phi} L} \frac{\Delta \Gamma(\frac{1}{2})\Gamma(\Delta +\frac{1}{2})}{\Gamma(\Delta) } +{\cal O}(G^2\gamma^4)\,,
\end{align}
 where $+/-$ respectively denotes the upper/lower singularities in the Penrose diagram.
Therefore as illustrated in  Figure~\ref{fig021}, the corresponding Penrose diagram is given roughly by a rectangular shape where the length of the horizontal side is 
larger than the vertical size. Hence the right side is causally further away from the left side and of course the L-R boundaries are causally disconnected 
from each other completely. 
As will be illustrated further below explicitly, the vev of operators $\langle O_\Delta (t) \rangle_R$ can be identified. Here and below, 
$L/R$  in the expectation value represents that the operator of interest is acting on  the left/right side Hilbert space. 
One finds in general 
that any initial perturbations of states 
will decay away exponentially in time. As was noted previously  for the other dimensions \cite{Bak:2017xla}, these solutions are describing thermalization of excited states above the thermal vacuum. This late-time exponential decay implies that the classical gravity description is inevitably coarse-grained \cite{Bak:2017xla}, whose nature in the context of the  AdS/CFT correspondence is explored in \cite{Bak:2017dkj}.

\subsection{$\Delta =1$ ($m^2=0$) case}\label{sec31}
The bulk  matter field  is dual to the operator of dimension $\Delta=1$.
Its solution reads
\bea
\chi = \gamma \cos \mu \cos  \tau\,,
\eea
where we set $\tau_C=0$ for simplicity.
Then the vev can be identified as
\bea
\langle O_1(t) \rangle_R = \frac{\gamma}{\cosh^2 \frac{2\pi}{\beta} t}\,, 
\eea
which is describing the thermalization with the dissipation time scale $t_d =\frac{\beta}{2\pi}$.
The metric remains to be  AdS$_2$ while the dilaton becomes
\bea
\phi= \bar\phi \tilde{L}\frac{(b+b^{-1})\cos \tau}{2\cos \mu}  -  8\pi G \gamma^2 \left[ 
\frac{1}{12} \cos^2 \mu \cos 2 \tau
   + \frac{1}{4} (1+ \mu \tan \mu) 
\right]\,.
\eea

\subsection{$\Delta =\frac{3}{2}$ ($m^2=\frac{3}{4}$) case}
The bulk  matter field  is dual to the operator of dimension $\Delta=\frac{3}{2}$.
Its solution reads
\bea
\chi = \gamma \cos^{\frac{3}{2}} \mu \cos \frac{3}{2} \tau\,,
\eea
where we set $\tau_C=0$ for simplicity.
Then the vev can be identified as
\bea
\langle O_{\frac{3}{2}} (t) \rangle_{R} = \frac{\gamma}{\cosh^{3} \frac{2\pi}{\beta} t} \left[
 \cosh \frac{\pi}{\beta} t -\sinh \frac{2\pi}{\beta} t   \sinh \frac{\pi}{\beta} t
\right]\,, 
\eea
which is describing the thermalization with the dissipation time scale $t_d =\frac{\beta}{2\pi}$.
The dilaton becomes
\bea
\phi= \bar\phi \tilde{L}\frac{(b+b^{-1})\cos \tau}{2\cos \mu} + 8\pi G \gamma^2 \left[ 
-\frac{3}{32} \cos^3 \mu \cos 3 \tau
+\frac{3}{8}\cos \mu -\frac{3}{4} \frac{1}{\cos \mu}
\right]\,.
\eea

\subsection{$\Delta =2$ ($m^2=2$) case}
This matter is dual to the scalar operator of dimension $\Delta =2$. 
The solution reads
\bea
\chi = \gamma \cos^2 \mu \cos 2\tau\,, 
\eea
by setting $\tau_C=0$ again. 
The vev then becomes
\bea
\langle O_2(t) \rangle_R =    \frac{\gamma }{\cosh^4 \frac{2\pi}{\beta} t}\left( 2- \cosh^2 \frac{2\pi}{\beta} t \right)\,,
\eea
which describes the thermalization of an initial excitation.
The dilaton becomes
\bea
\phi= \bar\phi \tilde{L}\frac{(b+b^{-1})\cos \tau}{2\cos \mu} + 8\pi G \gamma^2 \left[ 
-\frac{1}{10} \cos^4 \mu \cos 4 \tau+ \frac{1}{4} \cos^2 \mu-\frac{3}{4} (1+ \mu \tan \mu)
\right]. ~~~
\eea


\section{Janus two-sided black holes}\label{sec4}
In this section, we shall describe Janus deformations  \cite{Bak:2003jk,Bak:2007jm, Bak:2007qw}
of  AdS$_2$ 
black holes. 
These deformations basically make the Hamiltonians of the left-right boundaries differ from each other by turning on an exactly marginal operator\footnote{Deformed with an exactly marginal operator, each boundary system remains to be conformal. One may also consider Janus deformations by non marginal operators, which have an application in understanding quantum information metric of a given CFT \cite{Bak:2015jxd}.}.  The bulk scalar field $\chi $ with $m^2=0$ is dual to the exactly marginal operator denoted by $O_1$.

\subsection{Eternal Janus}
\begin{figure}
\vskip-1cm
\begin{center}
\includegraphics[width=6.3cm,clip]{janusnn.pdf}~~~~
\includegraphics[width=6.3cm,clip]{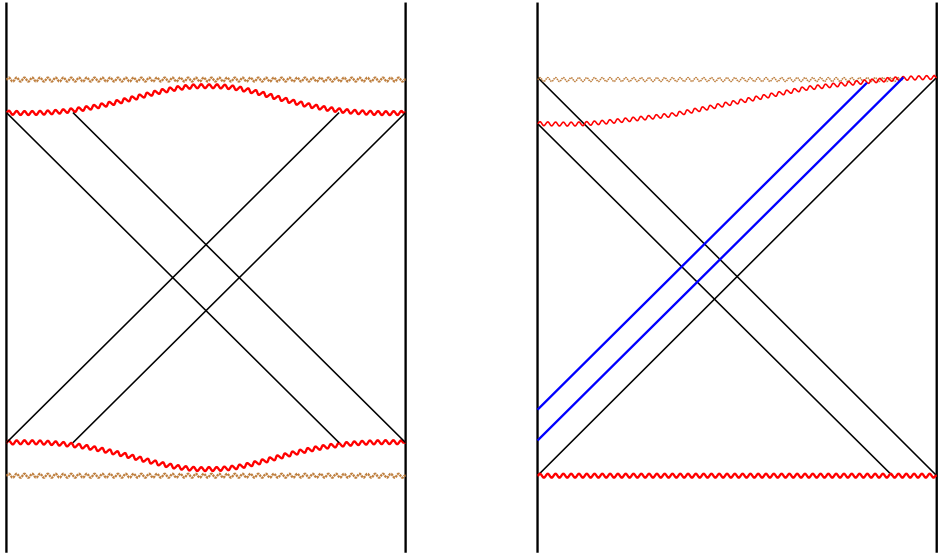}
\end{center}
\vskip-1cm
\caption{
\label{figst} \small Penrose diagram of the Janus deformed black hole is depicted on the left. On the right, we depict the deformation of AdS$_2$ black hole where only 
$H_l$ is deformed in a time dependent manner. 
}
\end{figure}

We begin with a simple Janus deformation given by
\bea
&&\chi = \gamma \left( \mu -\kappa_0\right)\,,  \nn
&& \phi= \bar\phi \tilde{L}\frac{(b+b^{-1})}{2}\frac{\cos \tau}{\cos \mu} -4\pi G \gamma^2  (1+ \mu \tan \mu)\,,
\eea
where $\tilde{L}$ is arbitrary and $b$ will be fixed as  a function of $\tilde{L}$ and the deformation parameter $\gamma$ later on. From the asymptotic values of 
the scalar field, we find that the L-R Lagrangians are deformed as 
\bea
{\cal L}_{L/R}( t_{L/R}) ={\cal L}_0(  t_{L/R}) + s_{L/R} (  t_{L/R}) O_1( t_{L/R})\,,
\eea
with source terms
\bea 
&&  s_L = - \gamma \left(\frac{\pi}{2}+\kappa_0\right)\,, \nn
&&  s_R = \ \ \gamma \left(\frac{\pi}{2}-\kappa_0\right)\,.
\eea 
Further using the standard dictionary of AdS/CFT correspondence, the vev of operators $O_1$ 
can be identified as
\bea
\langle O_{1}\rangle_{L/R} =\pm \frac{\gamma}{\cosh \frac{2\pi}{\beta} t}\,,
\label{janusvev}
\eea  
with the temperature given by
\bea
T= \frac{1}{2\pi} \frac{\tilde{L}}{ \ \ell^2}\,.
\eea
The dilaton part of the solution is left-right symmetric under the exchange of $\mu \leftrightarrow -\mu$. The parameter $b$ can be identified as
\bea
b - b^{-1} =\pm\frac {4 \pi^2 G }{\ \bar{\phi}\tilde{L}} \gamma^2\,,
\eea
where $-/+$ signature is the left/right side respectively. The corresponding Penrose diagram is depicted on the left of 
Figure \ref{figst} where the deformation of the singularity ($\phi=0$ trajectory) is denoted by wiggly red lines. It is clear that the L-R systems are still causally 
disconnected from each other completely.

Using the AdS/CFT dictionary developed in \cite{Bak:2007jm, Bak:2007qw}, the corresponding thermofield initial state can be constructed with a Euclidean evolution
\begin{equation} \label{}
{\cal U}_J = e^{-\frac{\beta}{4} H_r}  e^{-\frac{\beta}{4} H_l}\,,
\end{equation}
where $H_{l/r}$ denotes  the CFT Hamitonian obtained from ${\cal L}_{L/R}$ respectively. It is straightforward to check that the vev of the operator $O_1$, obtained from the field theory side  using the thermofield initial state and its time evolution, does indeed  agree with the gravity computation in (\ref{janusvev}).
Now by taking $\kappa_0 =\frac{\pi}{2}$, only $H_l$ on the left side is deformed while 
the Hamiltonian on the right side remains undeformed. We introduce the reduced density matrix 
\begin{equation} \label{}
\rho_R(t_R)={\rm tr}_L |\Psi (t_L, t_R)\rangle \langle \Psi (t_L,t_R) |\,,
\end{equation}
where we trace over the left-side Hilbert space. Without Janus deformation, one has $\rho_R(t)= \frac{1}{Z}e^{-\beta H}$, which is the usual 
time independent thermal density matrix. With the Janus deformation of $H_l$, one can view that the initial density matrix $\rho_R(0)$
is excited above the thermal vacuum.  The deformation also makes the left-right entanglement non-maximal. This excitation is relaxed away exponentially 
in late time, which is basically describing thermalization of  initial excitations. This explains the late time exponential decays of the vev
in (\ref{janusvev}). The relevant time scale here is the dissipation time scale $t_d = \frac{\beta}{2\pi}$.  Finally note that the scale $\tilde{L}$ is arbitrary.
One can view that this scale is set to be there from the outset $ t_{R/L}= -/+\infty$. Hence, although there is a time dependence in the vev,
 the system itself defined by $H_{l/r}$ is certainly time independent.  

\subsection{Excited  black holes}\label{ebh}
\begin{figure}[tb]
\begin{center}
	\includegraphics[width=10cm,trim={0 50mm 0 70mm},clip]{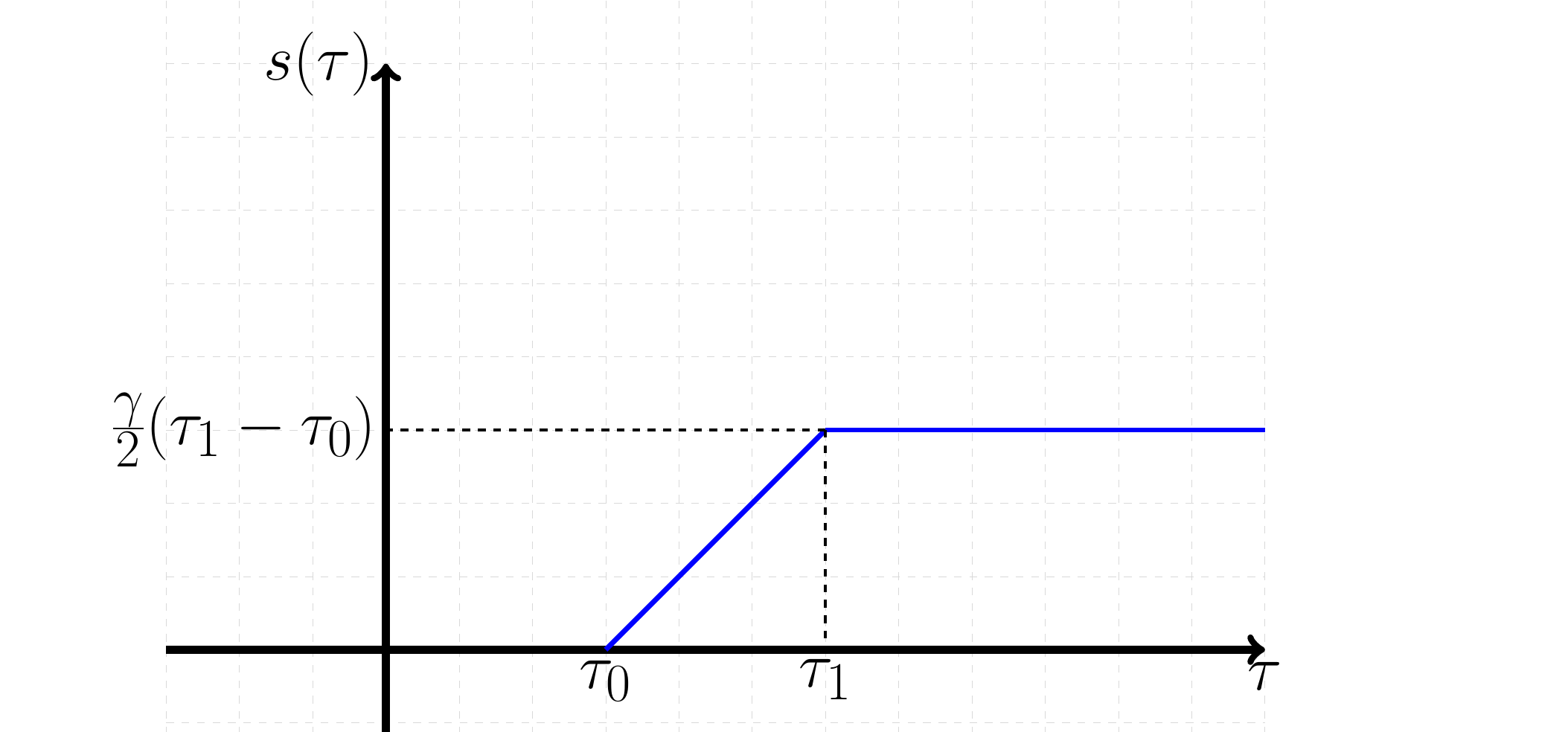}
\end{center}
\caption{
\label{source} \small  The source term of the operator $O_1$ as a function of $\tau$ is depicted. We turn on the deformation only for the left side
with $s_L =s(\tau)$.
}
\end{figure}

We would like to consider a solution that describes a change of system, 
beginning with a thermal state which is in equilibrium,  
and see its subsequent time evolution. To be definite, we shall only perturb the left side system at the moment by 
turning on the source term of an exactly marginal operator $O_1$ at some global time  $\tau=\tau_0$. In particular we shall consider 
a source term given by
\bea
s_L (\tau) =\left\{\begin{array}{cl}
0, & \ \ \ \ \    \, \tau < \tau_0 \,  \nn
\frac{\gamma}{2}(\tau-\tau_0),  & \ \ \ \ \   \, \tau_0 < \tau  < \tau_1 \, \nn
\frac{\gamma}{2}(\tau_1-\tau_0),  &  \ \ \ \ \   \, \tau_1 < \tau \, 
\end{array}
\right.
\eea
while $s_R(\tau)$ remains unperturbed in this example. We depict  the functional form of the source term in Figure \ref{source}. 
The corresponding scalar solution reads
\begin{equation} \label{}
\chi = \gamma (v- v_0)\Theta (v-v_0) -\gamma (v- v_1)\Theta (v- v_1)\,,
\end{equation}
where we introduce global null coordinates $u=\frac{1}{2}(\tau+\mu)$ and $v=\frac{1}{2}(\tau-\mu)$ and the initial/final 
values by $v_{0/1} =\frac{1}{2}\tau_{0/1}+\frac{\pi}{4} $.
As depicted on the right side of  Figure \ref{figst}, only the left boundary is initially deformed, whose effect is propagating through the bulk 
in a causally consistent manner.  The resulting bulk perturbation falls into the horizon and ends up with hitting future 
singularity which makes the dilaton part back-react. It never reaches the right-side boundary as dictated by the bulk causality. 
This is also clear from the field theory side since there is no interaction at all between the L-R systems and then 
no way to send any signal 
from one side to the other in this manner. 
To see the changes in the global structure of the spacetime, we turn to the description of the resulting changes in the dilaton field. 
 We shall denote its extra contribution produced by the  perturbation  by $\varphi$ and thus the full dilaton field becomes
\bea
\phi=\bar\phi L \frac{\cos\tau}{\cos \mu} +\varphi  \,.
\label{teleporteej}
\eea
Initially one has $\varphi=0$  until  $v$ becomes  $v_0$, so everything happens after $v_0$. For  $v_0 \le v \le v_1$,  the corresponding bulk stress tensor is nonvanishing
and 
 the resulting dilaton solution reads
\bea
\varphi = \frac{2\pi G \gamma^2}{ \cos \mu} \left[ \,\, 2(v-v_0) \sin \mu -\cos \mu + \cos (2v_0 -\tau)\,\,\right]\,.
\eea
For $v > v_1$, the bulk stress tensor vanishes and  the dilaton solution becomes
\bea
&&\varphi = \frac{4\pi G \gamma^2}{ \cos \mu} \left[ \,\,(v_1-v_0) \sin \mu \right.\,,
\nn
&& \  \  \  \left.  -\sin (v_1-v_0) \cos (v_1+v_0)\sin \tau + 
\sin (v_1-v_0) \sin (v_1+v_0)\cos\tau\,\, \right]\,,
\eea
which takes the form of the homogeneous one  in (\ref{hom}). Therefore the full solution in this region becomes the form of pure AdS$_2$ black hole described by
\begin{equation} \label{DilatonSol}
\phi= \phi_{BH}(\tilde{L}, b, \tau_B)
\end{equation}
where 
\begin{align}    \label{}
{\tilde{L}}^2 &= L^2 \Big[{\cal K}^2 -g_\gamma ^2 (v_1-v_0)^2\Big]\,, \nonumber \\
b^2 &= \frac{{\cal K}- g_\gamma (v_1-v_0)}{{\cal K}+g_\gamma  (v_1-v_0)}\,, \nonumber \\
\tan \tau_B &= -\frac{ g_\gamma  \sin (v_1-v_0) \cos (v_1+v_0)  }{1+ g_\gamma  \sin (v_1-v_0) \sin (v_1+v_0)  }\,.
\end{align}
Here, $g_{\gamma}$ is our perturbation parameter defined by $g_\gamma  =  \frac{4\pi G\gamma^2}{\bar\phi L} $ 
and  ${\cal K}$ denotes
\begin{equation} \label{}
{\cal K}^2= 1+ 2 g_\gamma  \sin (v_1-v_0) \sin (v_1+v_0)  + g_\gamma ^2  \sin^2 (v_1-v_0) \,.
\end{equation}
One finds $\tilde{L} \ge L$ at least to the leading order in $g_\gamma $ where we used the fact $0 \le v_0, v_1 \le \frac{\pi}{2} $. 
Hence the Beckenstein-Hawking temperature of the resulting black hole 
 increases after the perturbation. Though the scalar perturbation parameter $\gamma$ can take either signs, the dilaton perturbation parameter $g_0$ 
 is always non-negative definite.  Thus the change in temperature ought to be 
 independent of the signatures of the scalar perturbation parameter $\gamma$. This perhaps  reflects the fact the dual field theory side has to be
 strongly coupled 
 to have a gravity description.  
 In the gravity side, the black hole solution is independent of turning on a constant  moduli parameter
 described  by the constant part 
 of the scalar field. This  is consistent with the fact that our dilaton gravity description corresponds to the strong coupling limit of the boundary  quantum system.
 We shall not  explore any further details of the above aspect  in this note, since we are more concerned in the other aspects such as teleportation. 
 
 One finds $b\neq 1$ due to the perturbation and the shift parameter of the singularity is nonvanishing only in the left side. It can be identified as
\begin{equation} \label{}
 \Delta \tau_L = \tau_B - \arcsin \frac{g_\gamma (v_1-v_0)}{{\cal K}}\,,
\end{equation}
which is certainly  negative definite as expected. Thus we conclude that one cannot send a signal from one side to the other using the above perturbation 
since the L-R sides are causally disconnected with each other. This is quite consistent with the field theory side since no interaction between the left and the right boundaries
is turned on. We depict the resulting global structure on the right  of  Figure \ref{figst}.

Finally one may consider more general form of perturbation where the scalar solution takes the form
\begin{equation} \label{}
\chi =\chi_L =\sum^{n_L}_{i=0} \gamma^L_i \,  (v- v_i)\Theta (v-v_i)\,,
\end{equation}
where $\sum_i \gamma_i^L =0$ and we order $v_i$ such that $v_{i+1} > v_i$. Of course one may consider 
the perturbation
\begin{equation} \label{}
\chi =\chi_R =\sum^{n_L}_{i=0} \gamma^R_i \,  (u- u_i)\Theta (u-u_i)\,,
\end{equation}
where again  $\sum_i \gamma_i^R =0$ and we order $u_i$ such that $u_{i+1} > u_i$. This is describing the perturbation where the signal is 
sent from the right boundary. The only changes is we flip the L-R sides by the transformation $\mu \leftrightarrow -\mu$. The corresponding 
dilaton perturbations can be identified straightforwardly for the both types. When the both perturbations are present, interestingly one can get the corresponding solution
by a simple  linear superposition 
\begin{align}    \label{}
& \chi=\chi_L + \chi_R\,,\nonumber \\
& \varphi =\varphi_L +\varphi_R \,, 
\end{align}
where $\varphi_{L/R}$ is denoting the dilaton solution with the scalar field $\chi_{L/R}$ respectively\footnote{This superposition is possible only when one considers
the scalar field with $m^2=0$.}. Thus the solutions in the above describe rather general perturbations of the black hole system. We shall use these constructions to
describe  teleportees in later sections.

\section{Double trace deformation and stress tensor}\label{sec5}
In this section we consider the back-reaction in the dialton field by the  
1-loop stress tensor  of the  bulk scalar field  $\chi$ with the 
boundary condition which is related to the double trace deformation 
of the boundary theory. 
The bulk free scalar field $\chi$ in AdS space can have the asymptotic 
behavior $r^{-\Delta_{\pm}}$ along the radial direction $r$ where the 
fall-off power $\Delta_{\pm}$ is given by 
\eqref{deltapm}. 
When
$ -1/4<m^2<0 $, both modes of the power $\Delta_\pm$ become normalizable
and their duals may be realized as unitary scalar operators. In particular $\Delta_-$ 
is ranged over $(0,\frac12)$, which allows us to consider the double trace
deformation in the dual boundary theory \cite{Klebanov:1999tb,Witten:2001ua,Berkooz:2002ug}.  In the following we set
$\Delta \equiv \Delta_{-} $ and then $\Delta_{+} = d-\Delta$.
In this case, one may consider a general mixed boundary condition such 
that the boundary values of two modes become proportional. 

%
%
%

In our context for the coupling between the left and right boundary operators, the asymptotic behavior of the scalar field  in the right/left wedges in the Penrose diagram is given by
\begin{align}   \label{}
\chi(t,r)|_{R/ L} =  \frac{\alpha_{R/L}}{r^{\Delta}} + \cdots +   \frac{\beta_{R/ L}}{r^{1-\Delta}} + \cdots\,, 
\end{align}
and the mixed boundary condition corresponds to 
\begin{equation} \label{}
\beta_{L}(t) = h(-t)\alpha_{R}(-t)\,, \qquad \beta_{R}(t) = h(t)\alpha_{L}(-t)\,.
\end{equation}
According to the AdS/CFT correspondence for the double trace
deformation~\cite{Witten:2001ua}, this solution corresponds to the 
deformation of the Hamiltonian in the boundary theory given by
\begin{equation} \label{}
\delta H(t) = -h(t){\cal O}_{R}(t){\cal O}_{L}(-t)\,,
\label{defham}
\end{equation}
where $\mathcal{O}_{R,L}$ are scalar operators of dimension $\Delta$ dual to
$\chi$. This is a relavant deformation of dimension $2\Delta$ and the
coupling $h(t)$ has dimension $1-2\Delta$.

\begin{figure}[tb!] 
\begin{center}
\includegraphics[width=6.8cm]{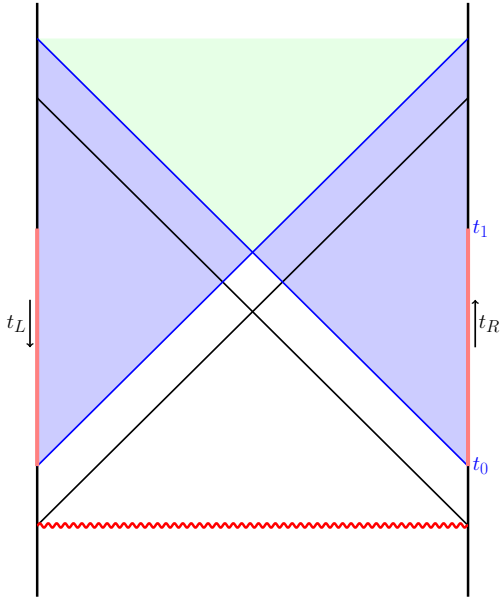}
\end{center}
\caption{The deformation $\delta H(t)$ is turned on at $t_{R/L}=\pm t_0$ and turned off
	at $t_{R/L}=\pm t_1$. The violet-colored right wedge is the intersection of 
	the spacelike region of the point $t_L=-t_0$ at the left boundary and 
	the timelike region of the point $t_R=t_0$ at the right boundary.
\label{figrw}
}
\end{figure}

Now suppose that $h(t)=0$ when $t<t_0$ so that we turn on the deformation 
at $t=t_0$. Then, as in the BTZ case~\cite{Banados:1992wn,Gao:2016bin},
the leading correction to the bulk 2-point function is expressed 
in the interaction picture as\footnote{This identification of $F$ and $T_{ab}$ below with the boundary interaction in (\ref{defham}) is 
correct only in a linearized order in $h(t)$
due to the problem of back-reaction effect in the identification of the boundary time. We shall take $F$ in (\ref{eq57}) and the resulting $T_{ab}$ in (\ref{Stress}) as our definition
of bulk interactions. Then we are left with the problem of identifying the corresponding boundary interactions including the higher order  effects. } 
\begin{align} \label{gh}
G  &\equiv F(t,t') + F(t',t) \notag \\
&=  i \int^{t}_{t_{0}}d\tilde{t}  \left\langle 
	  [\delta H(\tilde{t} ), \chi_{R}(t)], \chi_{R}(t')  \right \rangle
        + i \int^{t'}_{t_{0}}d\tilde{t}  \left\langle 
	  \chi_{R}(t) [\delta H(\tilde{t} ), \chi_{R}(t')]   \right \rangle\,,
\end{align}
where the coordinate $r$ is suppressed for simplicity. Here we are
interested in the violet-colored right wedge in Figure \ref{figrw}
which is the intersection of the spacelike region of the point $t_L=-t_0$ at
the left boundary and the timelike region of the point $t=t_R=t_0$ at
the right boundary. In particular, we
are not interested in the green-colored region deep inside the horizon
which is timelike from both boundary points at $t_{R/L}=\pm t_0$. Then
we can calculate $F(t,t')$ in the large $N$ limit by noting that 
$\mathcal{O}_L$ commutes with $\chi_R$  due to causality,
\begin{equation} \label{ftt}
F(t,t') \simeq -i \int_{t_0}^t d\tilde{t}  h(\tilde{t} ) 
\langle \chi_R(t') \mathcal{O}_L(-\tilde{t} ) \rangle
\langle [ \mathcal{O}_R(\tilde{t} ), \chi_R(t) ] \rangle\,.
\end{equation}
The first correlation function containing $\mathcal{O}_L$ may be
expressed in terms of the bulk-boundary propagator \eqref{BtoB} in
the appendix by use of the KMS condition~\cite{Martin:1959jp,Schwinger:1960qe,Mahanthappa:1962ex,Takahasi:1974zn} 
\begin{equation} \label{kms}
\langle \mathcal{O}_R(t) \mathcal{O}_L(t')\rangle_{\textrm{tfd}}
= \langle \mathcal{O}_R(t) \mathcal{O}_R(t'+i\beta/2)\rangle_{\textrm{tfd}}\,,
\end{equation}
and the second factor yields the retarded function \eqref{retarded}. Thus,
\begin{equation}
F(t,t') = 2 \sin\pi\Delta \int_{t_0}^t d\tilde{t} h(\tilde{t} ) K_\Delta (t'+\tilde{t} -i\beta/2)
                                            K_\Delta^r (t-\tilde{t} )\,.
                                            \label{eq57}
\end{equation}

The 1-loop stress tensor can be computed through the bulk 2-point function
$G(x,x')$ as 
\begin{equation} \label{Stress}
 T_{ab}   = \lim_{x'\rightarrow x}\Big[ \partial_{a}\partial'_{b}G (x,x')
 -\frac{1}{2}g_{ab}g^{\rho\sigma}\partial_{\rho}\partial'_{\sigma}G(x,x')
 - \frac{1}{2}g_{ab}m^{2}G(x,x')\Big]\,,
\end{equation}
where we have to subtract the singular expressions in the coincident limit.
As in the BTZ case~\cite{Gao:2016bin}, the zeroth order term in 
$h$ gives a vanishing contribution to $\int dU T_{UU}$ on the horizon $V=0$.
A nonvanishing result is obtained with the first order correction \eqref{gh}.
In the following, we assume that $h(t)$ is a nonvanishing constant 
only in the interval $[t_0, t_1]$,
\begin{equation}
h(t) = \begin{cases}
h (\frac{2\pi}{\beta})^{1-2\Delta},   &  \quad t_0 \le t \le t_1,  \\
	0, &  \quad    \text{otherwise} ~~\,,
\end{cases}
\end{equation}
where $h$ is constant. In Kruskal coordinates, we find
\begin{equation} \label{TUU}
T_{UU}(U,V) = 2\lim_{U',V'\rightarrow U,V}\partial_{U'}\partial_{U} F(U, U')\,,
\end{equation}
where $F(U,U')$ is given by
\begin{equation} \label{}
\label{FUU}
F(U, U')  =hN_{\Delta}\int^{U_f}_{U_{0}}\frac{d{\cal S}}{{\cal S}}\left[\frac{1+U'V'}{U'{\cal S}-V/{\cal S}+ 1 - U'V'}\right]^{\Delta}\left[\frac{1+UV}{U/{\cal S}-V{\cal S}-(1 - UV)}\right]^{\Delta}\,,
\end{equation}
with
\begin{equation} \label{}
N_{\Delta} \equiv \frac{1}{2\pi}~ \frac{2^{2\Delta-2}\Gamma^{3}(\Delta)}{\Gamma^{2}(2\Delta)\Gamma(1-\Delta)}\,,  \qquad  \qquad U_f   \equiv  %
\left\{ \begin{array}{ll}    
U\,,   &  \quad   \quad  U_0  \le  U <  U_1,  \\
U_1\,, &  \quad  \quad  ~~~ ~ U \ge  U_1 ~~~\,. \end{array}  \right.    \label{UfDef}
\end{equation}
%
%

%
%

\section{Traversable wormholes}\label{sec6}
In this section we compute the back-reacted deformation of the dilaton field through  the 1-loop stress tensor by the coupling between the left and right boundary operators. We will follow closely the steps in Ref.~\cite{Gao:2016bin} while some aspects could be addressed more concretely. One can see the reduction of the horizon radius and the uplift of the position of the singularity  compared before the deformation.  As a result, one can check that the thermodynamic 1st law holds in our setup, in addtion that wormholes become  traversable. Some formulae are relegated to Appendices for readability.

On the initial condition\footnote{Note that the boundary time $t_{i} (i=0,1)$ in the right wedge could be represented by $U_{i} = e^{Lt_{i}/\ell^{2}} = -1/V_{i}$ in the Kruskal coordinates. } of infalling matters such that $T_{UV}(U,V)=0$ for any value in the range of $U\le U_{0}$, the solution to (\ref{phieq}) of dilaton field $\phi$, sourced by the stress tensor $T_{ab}$, can be shown to become  (see  Appendix A) 
\bea \label{Defphi}
\phi &=& \bar\phi L \frac{\cos \tau}{\cos \mu}+ \varphi\,, \nn
\varphi &=&  - \frac{1}{1+UV}\int^{U}_{U_{0}}dP~  (U-P) (1+PV)T_{UU}(P,V)\,.
\eea
Note that, in terms of the global null coordinates with the change of a variable $P=\tan p$,
the dilaton field solution could also be written as 
\begin{equation} \label{}
\varphi(u,v) = \frac{1}{\cos(u-v)}\int^{u}_{u_{0}}dp~ \sin(p-u)\cos(p-v)~ T_{uu}(p,v)\,.
\end{equation}
By inserting the expression of $T_{UU}$ in 
(\ref{TUU}) to 
(\ref{Defphi}),  one can show, after some calculation,  that the deformation of the dilation field $\phi$   is given by (see Appendix B)
\begin{equation} \label{tworm}
\varphi(U,V) = 2h \Delta N_{\Delta} \int^{U_f}_{U_{0}}d{\cal S}\frac{{\cal S}^{2\Delta-1}}{(1+{\cal S}^{2})^{2\Delta}}\Big[ w^{1-\Delta}(1-w)^{2\Delta-1} + \Delta \frac{1+w}{1-w}B_{w}(1-\Delta,2\Delta)\Big]\,, 
\end{equation}
where $w$ is defined by
$$
w\equiv \frac{({\cal S}-V)(U-{\cal S})}{(1+{\cal S}V)(1+{\cal S}U)}\,,
$$
and $U_f$ is defined in Eq.~(\ref{UfDef}). 
Through the change of a variable $\tan s\equiv \frac{1}{2}({\cal S}-1/{\cal S})$, the above solution can also be written as
\begin{equation} \label{}
\varphi(u,v) = \frac{ h\Delta N_{\Delta}}{2^{2\Delta-1}}  \int^{s_{f}}_{s_{0}}ds\,  (\cos s)^{2\Delta-1}  \Big[ w^{1-\Delta}(1-w)^{2\Delta-1} + \Delta \frac{1+w}{1-w}B_{w}(1-\Delta,2\Delta)\Big]\,,
\end{equation}
where $w$ and $s_{f/0/1}$ are given  by 
\begin{equation} \label{}
w=\frac{\sin(u+v-s)-\cos(u-v)}{\sin(u+v-s)+\cos(u-v)}\,, \quad \tan s_{f} = \frac{1}{2}(U_f- \frac{1}{U_f})\,, \quad \tan s_{0/1}= \frac{1}{2}(U_{0/1} - \frac{1}{U_{0/1}})\,.  \notag
\end{equation}

\begin{figure}
\vskip-2cm
\begin{center}
\includegraphics[width=6.5cm]{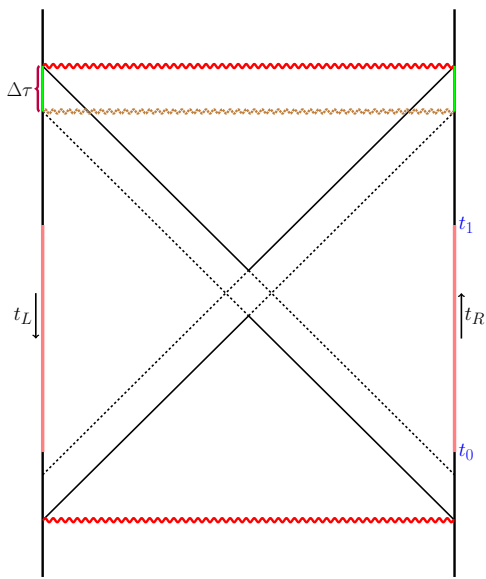}
\end{center}
\vskip-1cm
\caption{
\label{traw} \small Penrose diagram of the traversable wormhole is depicted in this figure. The left-right boundaries are now causally connected as the wormhole becomes traversable.
The shape of the upper singularity is not precise as it is curved in general.}
\end{figure}

Now, let us consider the near boundary region $\mu \simeq \pi/2$ after turning off the L-R coupling $h$, 
at which the singularity meets the right boundary of AdS$_{2}$. Note that this limit corresponds to $w=1$ in the incomplete Beta function $B_{w}(a,b)$. By using the expansion of $B_{w}(a,b)$ around $w=1$, the deformed dilaton field in the above limit becomes\footnote{Even before the L-R coupling turned off, this expression is valid once $\alpha_s$ and $\alpha_c$ below are replaced by those in (\ref{b40}) and (\ref{DilExp}). See Appendix \ref{appb} for the details.}
\begin{equation} \label{ExpDilaton}
\varphi(\tau,\mu) =  \frac{8\pi G}{\cos\mu} \Big[h \alpha_{s} \sin \tau   + h \alpha_{c}\cos \tau\Big] +O(\cos^{2\Delta}\mu)\,,
\end{equation}
where $\alpha_{s}$ and $\alpha_{c}$ are given by
\begin{align}   \label{}
 \alpha_{s} &=  2\Delta^{2} N_{\Delta}B(1-\Delta,2\Delta)~ \left[ B_{z_0}(\Delta+\frac{1}{2},\Delta+\frac{1}{2})-
B_{z_1}(\Delta+\frac{1}{2},\Delta+\frac{1}{2})\right]\,,
 \nn 
 \alpha_{c} &=  -\Delta N_{\Delta}B(1-\Delta,2\Delta)~ \Big[ z_0^{\Delta}(1-z_0)^{\Delta}-z_1^{\Delta}(1-z_1)^{\Delta}\Big]\,,   \quad \quad z_i \equiv \frac{1}{1+U^{2}_{i}}\,.  \qquad 
\end{align}
Combined with the homogeneous part of the solution, the full solution becomes 
\begin{equation} \label{sol67}
\phi(\tau,\mu) = \bar\phi L \frac{\cos\tau}{\cos \mu}+  \frac{8\pi G}{\cos\mu} \Big[h\alpha_{s}\sin \tau    + h\alpha_{c}\cos \tau\Big] +O(\cos^{2\Delta}\mu)\,.
\end{equation}
%
We would like to understand the deformation of this geometry and the structure of the singularity in the asymptotic region of our interest.  We compare  this with the 
general form of the static black hole solution 
\begin{equation} \label{}
\phi=\phi_{BH}(L+\delta L, 1+\delta b, \delta \tau_B ) \,,
\end{equation}
where $\delta L$, $\delta b$ and $\delta \tau_B$ are describing 
deformation of the black hole due to the L-R coupling. 
This leads to the identification of parameters 
\begin{align}    \label{}
\left(1+\frac{\delta L}{L}\right)^2 &= \left(1+\frac{8\pi G}{\bar{\phi} L} h\alpha_c\right)^2 +  \left(\frac{8\pi G}{\bar{\phi} L} h\alpha_s\right)^2\,,   \nonumber \\
\delta \tau_B  &= \arctan \frac{\frac{8\pi G}{\bar\phi L} h\alpha_s}{1+\frac{8\pi G}{\bar{\phi} L} h\alpha_c}\,,   \nonumber \\
\delta b &= 0\,.
\end{align}
These parameters  could also be read from (\ref{kru}) and (\ref{BackDil}) in Kruskal coordinates directly which we have done to the linear order in $G h$. 
Since the static black hole is characterized by the temperature only as we discussed in section \ref{sec2}, one relevant physical parameter is $\delta L$ that is related to 
the temperature $T +\delta T$ with $\delta T= \frac{1}{2\pi}\frac{\delta L}{\ell^2}$.  The other relevant physical parameter is the shift of singularity identified as
\bea
\Delta \tau^R_+ =  \delta \tau_B \sim \frac{8\pi G}{ \bar{\phi} L}  h \alpha_s +O(G^2 h^2)\,.
\eea  
This expression 
may take either signs.  We shall choose $h$ positive where the shift becomes 
positive-definite.
It then tells us that the wormhole becomes traversable since the position of the singularity is uplifted in the right wedge as can be seen from Figure~\ref{traw} by the amount $\Delta \tau^R_+$. 
Since our configuration is left-right symmetric under the exchange of $\mu \leftrightarrow -\mu$, one has $\Delta \tau^R_+ =\Delta \tau^L_+$ both of 
which are denoted as $\Delta \tau$ in Figure~\ref{traw}.
Note also that the expression of $\alpha_{s}$ is effectively identical to the expression of the averaged null energy condition violation $\int^{\infty}_{U_{0}}dU\, T_{UU}$. In this manner, one can send a signal from one side to the other through the wormhole now.
This wormhole parameter is monotonically increasing as a function of $z_0 \in [0,1]$. This implies that the earlier the L-R interaction starts, 
the bigger the traversable gap opens up, which is in accordance with our physical intuition. Its maximum is attained at $z_{0}/z_1=1/0$ corresponding to $U_{0}/U_1=0/\infty$
or $t_{0}/t_1 =-/+\infty$. 

In our two-dimensional model of black holes, the energy $E_{R}$ could be identified with the black hole mass $M$, and the black hole entropy change can be read from the change of horizon area as
\begin{equation} \label{}
\delta S_{BH} = \frac{\bar\phi}{4G}\delta L={\cal C}\delta T\,
\end{equation}
to the leading order in $Gh$ taking the small variation limit.
%
%
%
Note that, once $t_0 \ge 0$, the horizon radius is reduced by the amount $\delta L =  \frac{8\pi G}{\bar{\phi} L} \alpha_c < 0$ and the Beckenstein-Hawking temperature
will be reduced accordingly as well.   
This reduction of the horizon radius is consistent with the entropy reduction in the measurement process during the quantum teleportation, which is argued to be dual to the traversable wormhole. 

To see the consistency of this change of the horizon radius, let us return to the energy change by the deformed Hamiltonian in the boundary theory. Using expression for 
the black hole energy (or mass) in (\ref{energybh}), the change in energy is identified as 
\begin{equation} \label{}
\delta E={\cal C} T \delta T = \frac{L }{\ell^2} h \alpha_c\,.
\end{equation}
In fact this expression can be confirmed directly from the boundary computation as follows. 
Note that, along the method in the Ref.~\cite{Gao:2016bin}, the change of energy by the Hamiltonian deformation is given by 
\begin{align}   \label{kkk}
\delta E_{R} & =-i \int^{t_{1}}_{t_{0}}dt' \langle \Psi(t')| [\delta H(t'),H_{R}]| \Psi(t') \rangle     \nn 
 &= \int^{t_{1}}_{t_{0}}dt'h(t') \langle \Psi(t') |\dot{\cal O}_{R}{\cal O}_{L}(t') | \Psi(t')\rangle    \nn
&=  h\Big(\frac{2\pi}{\beta}\Big)^{1-2\Delta}\int^{t_{1}}_{t_{0}}dt'~ \frac{1}{2}\partial_{t'}\Big[\lim_{r'\rightarrow\infty}\Big(\frac{r'}{\ell^{2}}\Big)^{\Delta}K_{\Delta}(2t'-\frac{i}{2}\beta) \Big]  \nn
&= h\Big(\frac{L}{\ell^{2}}\Big) \frac{\Gamma^{2}(\Delta)}{8\pi\Gamma(2\Delta)}\Big[\frac{1}{(\cosh\frac{L t_{1}}{\ell^{2}})^{2\Delta}}- \frac{1}{(\cosh\frac{L t_{0}}{\ell^{2}})^{2\Delta}}\Big]\,,
\end{align}
where we use the undeformed thermofield  initial state defined in (\ref{initial}) with ${\cal U}={\cal U}_0$.
%
%
%
Further noting $U_{i} = e^{Lt_{i}/\ell^{2}}$ in the right boundary,
%
it is then obvious that  the first law holds in our limit 
\begin{equation} \label{}
\delta E_{R}= \delta M = T  \delta S \,. 
\end{equation}
%

We find that   $\delta E_R$ or $\delta S$   becomes positive if $0 \ge  t_1 >  t_0$ but 
 the wormhole is still traversable.     
This behavior is related to  thermalization of perturbation as suggested by the time dependence in (\ref{kkk}) shows exponential decaying behaviours with the dissipation time 
scale given by $t_d =\frac{\beta}{2\pi}$.  
This earlier perturbation still makes wormhole traversable while the perturbation is allowed to be thermalized enough to have an increased total entropy 
of the system. Of course by the first law, the corresponding change of energy should be positive.  The overall picture here is nothing new. We just make its bulk 
description direct and quantitative.

Finally one may  show that, to the leading in $h$, our traversable wormhole solution in this section is indeed consistent with that from the boundary reparametrization dynamics in \cite{Maldacena:2017axo,Maldacena:2018lmt}. See Appendix \ref{eee} for the detailed comparison.


\section{Full bulk teleportation}\label{sec7}
In this section, we shall present a simple teleportation model in the boundary side and various bulk solutions which include
 a traversable wormhole and a teleportee state traveling through the wormhole from the left to the right
boundaries.    

Teleportation is sending a quantum state to a remote place via EPR entanglement channel. Let us give its elementary introduction here.  Alice on the left side would like to teleport
 a qubit
\begin{equation} \label{}
|T\rangle =c_0 |0\rangle_T + c_1 |1 \rangle_T \,, 
\end{equation}
from the left to the right boundaries. We model the L-R entanglement by an EPR pair
\begin{equation} \label{}
|\Psi \rangle_{LR} = \frac{1}{\sqrt{2}}\sum^1_{i=0}   | i \rangle_L | i \rangle_R
\,,
\end{equation}
where $i,j,k,\theta_1,\theta_2=0,1$.  This $|T\rangle$ is added to the left side at some point. One then represents the left side system in a new basis spanned by 
\begin{equation} \label{}
|\theta_1 \theta_2 \rangle_{M} =\frac{1}{\sqrt{2}}\left[ \, |\theta_1 0\rangle_{TL} + (-)^{\theta_2} |1-\theta_1~~  1 \rangle_{TL} \right]\,,
\end{equation}
each of which is maximally entangled. With this new basis, the full system can be represented by
\begin{equation} \label{}
|T\rangle |\Psi \rangle_{LR} = \frac{1}{2}\sum_{\theta_1,\theta_2,i,k} |\theta_1 \theta_2 \rangle_{M} (U^{-1}_{\{\theta_1\theta_2\}})_{ik}c_k  | i \rangle_R\,,
\end{equation}
where $U_{\{\theta_1\theta_2\}}= \sigma^{\theta_1}_1 \sigma^{\theta_2}_3$ with $\sigma_{1,2,3}$ denoting the standard Pauli matrices.
Now Alice 
makes a measurement in the $M$ basis ending up with a particular state $|\theta_1 \theta_2 \rangle_{M}$ and sends its result
$\{\theta_1, \theta_2\}$ to Bob on the right side via an extra classical channel. This classical channel is completely independent of our bulk. 
At this stage the total state becomes 
\begin{equation} \label{}
 \sum_{i,k} |\theta_1 \theta_2 \rangle_{M} (U^{-1}_{\{\theta_1\theta_2\}})_{ik}c_k  | i \rangle_R\,.
\end{equation}
Once Bob gets the message, he performs a unitary transform of his state by the action $V_\theta =U_{\{\theta_1\theta_2\}}$ and then  the resulting  state becomes
\begin{equation} \label{}
 |\theta_1 \theta_2 \rangle_{M}  \sum_{k}  c_k  | k \rangle_R\,.
\end{equation}
This completes the quantum teleportation of $|T\rangle $ from the left to the right. Of course one can consider more general setup  where one has an L-R entanglement involving  many qubits
and teleports  more than one qubit.   In particular when one uses  our Einstein-Rosen bridge as the L-R entanglement based on the so called ER=EPR relation \cite{Maldacena:2013xja},  there will be in general a thermailzation of $|T \rangle $ state after 
its inclusion to the left side. Measurement can be made by picking up an arbitrary qubit of $L$ system and forming a maximally  entangled basis where we assume $|T\rangle$ is one qubit. After the measurement, Alice again sends 
the result of measurement to Bob. Bob then recovers the    $|T \rangle $ by the action of an appropriate unitary transformation.  For more detailed discussions, we refer  
to Ref.~\cite{Susskind:2017nto}. 

A few comments are in order. First of all, the measurement in general makes the L-R entanglement reduced and  
the L-T system entangled instead. The second essential feature is the L-R 
coupling by the measurement $M^\theta_L$ on the left side and the recovery action $V^\theta_R$ on the right side. This coupling basically makes the wormhole traversable as we verified
in the previous section.

We now turn to our bulk description. We shall basically combine the traversable wormhole 
in (\ref{tworm}) and the bulk-traveling solution (\ref{teleporteej}).  
As was done in Ref.~\cite{Maldacena:2017axo}, we would like to suppress any higher loop corrections in our computation. For this purpose, we shall introduce 
$K_\theta$ dimension $\Delta$ 
and $K_T$ dimension one operators respectively for the L-R coupling of the traversable wormhole and the teleportee 
degrees and excite them altogether coherently. We shall fix parameters ${\bar{\gamma}}^2$ and $\bar{h}$ which are defined by
\begin{align}    \label{}
{\bar{\gamma}}^2 &= \frac{4\pi G }{\bar\phi L}K_T \gamma^2 \,, \nonumber \\
{\bar{h}}\  &= \frac{8\pi G }{\bar\phi L} K_\theta h\,,
\end{align}
while taking large  $K_{T,\theta}$  limit. This then makes any possible higher loop corrections  suppressed. The combined solutions can be presented
in the form 
\begin{equation} \label{}
\phi  =  \frac{\bar\phi L}{\cos \mu} \left[\, Q_\mu \sin\mu + Q_s \sin \tau + 
(1+  Q_c) \cos \tau\,\right] + O(\cos^{2\Delta}\mu)\,,
\end{equation}
in near right-boundary region where the L-R coupling and boundary interactions are not present or were already turned off.  The wormhole opening parameter of the right boundary
 is given by
 \be
 \Delta \tau^+_R = \arctan \frac{Q_s}{1+Q_c}
 \ee
 and the traversability condition of the resulting 
wormhole requires $Q_s > 0$ which leads to positive $\Delta \tau^+_R$. Let us begin with a minimal one. 

Solution $A$ :  For $v_0 > v \ge 0$ and $u \ge  \frac{1}{2} \left(s_1+\frac{\pi}{2}\right) $\footnote{Recall that our L-R interaction lasts for $s_1> s > s_0 $, where $s$ is the boundary global time coordinate.}, 
one has 
\begin{equation} \label{} 
Q_s =\bar{h} \alpha_s\,,  \qquad  Q_c =\bar{h} \alpha_c\,,
\end{equation}
with $Q_\mu=0$.
On the other hand, for $v \ge v_1$ and $u\ge  \frac{1}{2} \left(s_1+\frac{\pi}{2}\right)$, 
\begin{align}    \label{}
& Q_\mu =   {\bar{\gamma}}^2 (v_1-v_0)\,, \nonumber \\
& Q_s =\bar{h} \alpha_s- {\bar{\gamma}}^2 \sin (v_1-v_0)\cos (v_1+v_0)\,,  \nonumber \\
&  Q_c =\bar{h} \alpha_c +{\bar{\gamma}}^2 \sin (v_1-v_0)\sin (v_1+v_0)\,.
\end{align}
Since $\bar{h}\alpha_s$ is positive definite with our choice $h >0$, the wormhole becomes traversable. Then one may send a signal from the left to the right
while the wormhole is traversable. Therefore the $|T\rangle $ state added to the left side will appear on the right side which is the teleportation from the left to the 
right boundaries. One finds that the wormhole opening parameter $\Delta \tau^+_R$ is dependent in general on the degrees and details  of the 
teleportee. We would like to take $v_1 \le \frac{\Delta \tau^+_R}{2}$ such that no information is lost into the singularity behind the horizon. Since the added contribution 
to $Q_c$ by the teleportee is given by
\begin{equation} \label{}
{\bar{\gamma}}^2 \sin (v_1-v_0)\sin (v_1+v_0)\,.
\end{equation}
which is always positive definite. Then there is a corresponding increase of the black hole entropy and mass, which may be used as a criterion about how many qubits are
transported\footnote{There is also a (negative) contribution of entropy due to the (anti) themalization effect.}.

Let us now turn to the case where Bob also throws a matter into the horizon from the right. In our case, this matter wave  is again described by the massless  scalar field  introduced
 in Section \ref{ebh}. Namely,
we consider the left-moving wave
\bea
\chi_R = \gamma_R (u- u_0)\Theta (u-u_0) -\gamma_R (u- u_1)\Theta (u- u_1) 
\eea
in addition to the L-R coupling and the teleportee of Solution A. Here of course we take $u_1 \ge u_0$ with $u_i \in [0, \frac{\pi}{2} ]$. The resulting full system
 is described by the following.

Solution B: For the region $v_0 \ge v \ge   0 $ 
 and $u \ge {\rm max}\left(u_1, \frac{1}{2} \left(s_1+\frac{\pi}{2}\right)\right)$, 
the dilaton is described by
\begin{align}    \label{sol714}
& Q_\mu =  - {\bar{\gamma}_R}^2 (u_1-u_0)\,, \nonumber \\
& Q_s =\bar{h} \alpha_s- {\bar{\gamma}}_R^2 \sin (u_1-u_0)\cos (u_1+u_0)\,,  \nonumber \\
&  Q_c =\bar{h} \alpha_c +{\bar{\gamma}}_R^2 \sin (u_1-u_0)\sin (u_1+u_0)\,.
\end{align}
For the region  $v \ge v_1$ and $u \ge {\rm max}\left(u_1, \frac{1}{2} \left(s_1+\frac{\pi}{2}\right)\right) $, 
 the dilaton parameters are given by
\begin{align}    \label{}
& Q_\mu ={\bar{\gamma}}^2 (v_1-v_0)  - {\bar{\gamma}_R}^2 (u_1-u_0)\,,   \nonumber \\
& Q_s =\bar{h} \alpha_s- {\bar{\gamma}}^2 \sin (v_1-v_0)\cos (v_1+v_0)- {\bar{\gamma}}_R^2 \sin (u_1-u_0)\cos (u_1+u_0)\,,  \nonumber \\
&  Q_c =\bar{h} \alpha_c +{\bar{\gamma}}^2 \sin (v_1-v_0)\sin (v_1+v_0)
+{\bar{\gamma}}_R^2 \sin (u_1-u_0)\sin (u_1+u_0)\,.
\end{align}
%

\begin{figure}[tb!] 
\vskip-0.8cm
\begin{center}
\includegraphics[width=6.3cm]{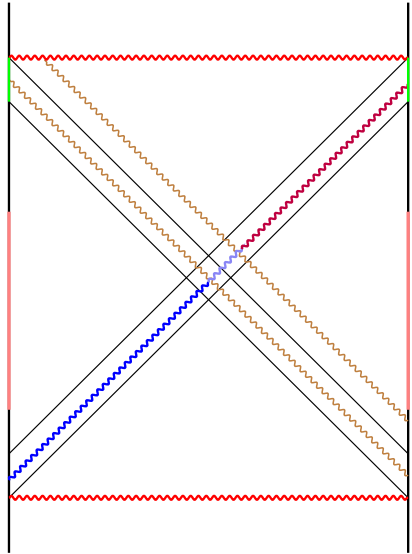}
\end{center}
\vskip-0.8cm
\caption{ The full bulk teleportation with additional matters thrown into the horizon from the right side. The teleportee will be affected by this bulk encounter.
Again the shape of the upper singularity is not precise as it may be curved in general.
\label{figfwh}
}
\end{figure}

When the perturbation from the right side is too strong such that 
\begin{equation}   
{\bar{\gamma}}_R^2 \sin (u_1-u_0)\cos (u_1+u_0)      \ge \bar{h} \alpha_s\,,
   \label{condition}
\end{equation}
the wormhole is no longer traversable since $\Delta \tau^+_R$ of the solution in (\ref{sol714}) becomes non positive. Of course, one has  then a usual two-sided black hole system
where any perturbations thrown into horizon hit the future singularity inevitably. 
If the extra matter thrown into the black hole from the right side is not too big such that the condition (\ref{condition}) is violated, 
the wormhole remains traversable and the teleportee
can be sent from the left to the right sides through the wormhole.  It is clear that the teleportee will meet the matter from the right side 
while transported. It can record and  report this encounter to Bob on the right side. Hence one may conclude that the bulk 
wormhole is experimentally observable, which was emphasized in Ref.~\cite{Susskind:2017nto}. 

As was mentioned in the introduction, in this note we have ignored any possible back-reaction effect caused by the bulk excitations. Once there are any excitations, identification of the time 
coordinate $t_{L/R}$ of the boundary system as a function of the bulk time coordinate $t$ becomes nontrivial. Without any excitations, one has an 
identification $ t(t_{L/R})=t_{L/R}$ for the AdS black holes of Section  \ref{sec2}. With excitations, the reparametrization dynamics $t(t_{L/R})$ can be nontrivial,
which is one of the main points of  Refs.~\cite{Maldacena:2016upp, Maldacena:2017axo}.  Indeed the presence of the bulk teleportee will make the 
$t(t_{L/R})$ dynamically nontrivial. Hence this back-reaction effect has to be taken into account\footnote{Our bulk solutions are fully valid in their own right. 
However, without full identification of $t(t_{L/R})$, we simply do not have their precise boundary interpretation. }. 
In order to fix the action of the L-R interactions in the 
presence of this back-reaction effect, one then has to modify the L-R coupling strength $h$ in an appropriate manner. 
For a given L-R interaction specified by the boundary time $t_{L/R}$, one may ask if there is a limit in the number of qubits that can be teleported \cite{Maldacena:2017axo}.
In our formulation, we are not able to show this limit which requires the identification of $t(t_{L/R})$. 

Alternatively one may ask the following 
to see the above limitation in our formulation. Namely one may effectively makes 
$\bar{h}$ together 
with the number of qubits of the teleportee state small enough such that the above back-reaction effect is negligible. 
 In this situation, one might  be able to see the above limitation. For this, we take $\bar{h} \ll 1$. Then clearly $\Delta \tau^R_+$ 
 is of $O(\bar{h})$. Then $v_0, v_1 \sim  O(\bar{h})$ since $v_1 \le \frac{1}{2}\Delta \tau^R_+$. 
 The resulting change in the entropy due to the presence of the teleportee
 will be of $O(\bar{\gamma}^2 {\bar{h}}^2)$ since 
\begin{equation} \label{}
\Delta S \propto {\bar{\gamma}}^2 \sin (v_1-v_0)\sin (v_1+v_0) \sim O(\bar{\gamma}^2 {\bar{h}}^2)\,.
\end{equation}
%
Since Solution A has the dependence on the parameter ${\bar{\gamma}}^2  (v_1-v_0)$, which we would like to keep as small as $O(\bar{h})$ to control the approximation.  
(The entropy change  is  the order of the back-reaction effect $O({\bar{h}}^2)$ which we have ignored in our approximation.) At any rate, the teleported bits are too small
and it is not possible to see any limitation in this manner. 
Further study in this direction is required.

\section{Conclusions}\label{sec8}
In this note, we have described a complete bulk picture of the quantum 
teleportation through a traversable wormhole 
in the two-dimensional dilaton gravity with a scalar field.
First, we have constructed various full back-reacted
solutions describing perturbation of a black hole and its thermalization. 
To realize a teleportee state, we have considered a specific 
time-dependent Janus deformation of AdS$_2$ black hole where only the left
boundary is initially deformed but the effect is propagating through the
bulk obeying the causality. This solution by itself cannot be used
to send signal from one boundary to the other because the dilaton back 
reacts in the way that the signal hits the future singularity before 
reaching the other side of the boundary. This is consistent with the field
theory side since L-R systems are completely decoupled from each other and
hence there is no way to send any signal from one side to the other.

The situation, however, changes if we turn on the double trace interaction 
between two boundaries which violates the average null-energy condition in 
the bulk. It renders the wormhole traversable and, to the leading order,
the traversability can be maintained in the presence of the teleportee 
state. We have solved of the equation of motion for the dilaton
for a general stress tensor. It allows us to identify
the relevant parameters which are responsible for
the transversability and the change of the horizon area. The entropy
is changed consistently with the black hole first law.
Our solutions have then further be extended to include extra matter thrown 
into the the black hole from the boundary that the teleportee would reach. 
The teleportee would meet the matter during transportation, which shows that
the bulk wormhole is experimentally observable.

In this note, we have not attempted to include any possible back reactions in
identification of the boundary time coordinates $t_{L/R}$ as a function of 
the bulk time coordinate $t$. Though our bulk solutions are fully 
consistent in their own right, we need a precise identification of
$t(t_{L/R})$ in order to have their proper boundary interpretation. Indeed
it is clear that the naive identification $t(t_{L/R})=t_{L/R}$
for the AdS black hole should be modified in the presence of the 
bulk teleportee and other bulk interactions. This would require the formulation of 
Ref.~\cite{Maldacena:2016upp} and further study is needed in this direction.


\section*{Acknowledgement}
D.B. was
supported in part by
NRF Grant 2017R1A2B4003095, and by Basic Science Research Program
through National Research Foundation funded by the Ministry of Education
(2018R1A6A1A06024977).
S.-H.Y. was supported by the National Research Foundation of Korea(NRF) grant with the grant number NRF-2015R1D1A1A09057057.

\appendix

\section{Coordinates and dilaton solution}
In this appendix, we would like to solve the equation of motion \eqref{phieq}
in the presence of the energy-momentum tensor $T_{ab}$. In the Kruskal
coordinates of AdS$_2$, the metric is given by
\begin{equation}
	ds^2 = -\frac{4\ell^{2}dUdV}{(1+UV)^{2}}\,.
\end{equation}
The Kruskal and global coordiantes for AdS$_{2}$ are related by
\begin{equation} \label{}
U=\tan \frac{\tau+\mu}{2}\,, \qquad V=\tan\frac{\tau-\mu}{2}\,,
\end{equation}
and in the Kruskal  coordinates, \eqref{phieq} becomes
\begin{align} \label{eqphiK}
\frac1{(1+UV)^2} \partial_U \left[ (1+UV)^2 \partial_U \phi \right] &=
	-8\pi G T_{UU} \notag \\
\frac1{(1+UV)^2} \partial_V \left[ (1+UV)^2 \partial_V \phi \right] &=
	-8\pi G T_{VV} \notag \\
-\partial_U \partial_V \phi - \frac2{(1+UV)^2} \phi &= -8\pi G T_{UV}
\end{align}
where 
the energy-momentum tensor $T_{ab}$ satisfies the conservation
$\nabla^a T_{ab}=0$. Explicitly,
\begin{align} \label{tabcons}
(1+UV)^2 \partial_V T_{UU} + \partial_U\left[(1+UV)^2 T_{UV}\right] &= 0 \,,
	\notag \\	
(1+UV)^2 \partial_U T_{VV} + \partial_V\left[(1+UV)^2 T_{UV}\right] &= 0  \,.
\end{align}
\eqref{eqphiK} can readily be integrated to obtain the general solution
of $\phi(U,V)$. In the following we will set $8\pi G =1$. Due to the conservation equation \eqref{tabcons}, there
are several different ways to express the solution. One way is to 
express it in a symmetric way with respect to $U$ and $V$,
\begin{align}
\phi(U,V) = - \frac1{1+UV} & \left[ \int_{U_0}^U (U-t)(1+tV)T_{UU}(t,V_0)\, dt
  + \int_{V_0}^V (V-s) (1+sU)T_{VV}(U_0,s)\, ds \right. ~~~ \notag \\
  &\left. - \int_{V_0}^V \int_{U_0}^U \frac{2(tV+sU)+(1-UV)(1-st)}{1+st}
		       T_{UV}(t,s)\, dt\,ds \right] \,,
\end{align}
where we have omitted the homogeneous solution $\phi_{hom}$ without the 
source $T_{ab}$. This homogeneous solution  may be given, for constants $\alpha_{i}$,  as 
\begin{equation} \label{Homsol}
\phi_{hom}(U,V) =  \alpha_{0} \, \frac{1-UV}{1+UV} + \alpha_{1}\, \frac{U+V}{1+UV} + \alpha_{2}\, \frac{U-V}{1+UV}\,,
\end{equation}
which is the same expression in  (\ref{hom}) rewritten in the Kruskal coordinates. 
Note that two of constants $\alpha_{i}$ could be set to zero by using the
$SL(2,{\bf R})$ isometry of AdS$_{2}$ space. In the following, we will set
$\alpha_{0}=\bar{\phi}L$, $\alpha_{1}=\alpha_{2}=0$, which corresponds to $b=1$ and $\tau_{B}=0$ in (\ref{DilatonSol0}) and $\bar{\phi}$ is taken as unity, $\bar{\phi}=1$, just for brevity.  It is straightforward to check that this solution satisfies \eqref{eqphiK}.

Now, we present an asymmetric form of the dilation solution, which contains a single integral. 
Note that a component of the conservation equation $\nabla^a T_{ab}=0$ is given by 
\begin{equation} \label{}
\partial_{s}T_{UU}(t,s) + \partial_{t}T_{UV}(t,s)  + \frac{2s}{1+ts} T_{UV}(t,s) =0\,,
\end{equation}
whose integration leads to 
\begin{equation} \label{}
T_{UU}(t,V_{0}) - T_{UU}(t,V)   = \int^{V}_{V_{0}}ds\Big[\partial_{t}T_{UV}(t,s) + \frac{2s}{1+ts}T_{UV}(t,s)\Big]\,.
\end{equation}
By multiplying $(U-t)(1+tV)$, one can see that 
\begin{align}   \label{}
& (U-t)(1+tV)\Big[ T_{UU}(t,V_{0}) - T_{UU}(t,V) \Big]  \nn
=   & \int^{V}_{V_{0}}ds~  \left\{ \partial_{t} [(U-t)(1+tV)T_{UV}(t,s)]
+ \frac{ 2(tV+sU) + (1-UV)(1-st) }{1+ts}T_{UV}(t,s) \right\}\,.  \nonumber
\end{align}
Inserting the above expression of $(U-t)(1+tV) T_{UU}(t,V_{0})$ into the previous symmetric form  of the dilaton  solution and then integrating over $t$-variable, one can see that the dilaton solution is  also written in a single integral expression as follows:
\begin{align}   \label{DSolS}
\phi(U,V) =  - \frac1{1+UV} & \left[ 
\int^{U}_{U_{0}}dt (U-t)(1+tV) T_{UU}(t,V) 
+\int^{V}_{V_{0}}ds(V-s)(1+sU)  T_{VV}(U_{0},s) \right. \nn 
& \left. - \int^{V}_{V_{0}}ds(U-U_{0})(1+U_{0}V) T_{UV}(U_{0},s) \right] \,.
\end{align}
Note that,  by taking the initial condition such that $T_{ab}(U,V)=0$ for any $U\le U_{0}$, the solution of dilaton field $\phi$ in (\ref{DSolS}) reduces to
\begin{equation} \label{phiTexp}
\phi(U,V) =  - \frac{1}{1+UV}\int^{U}_{U_{0}}dP~  (U-P) (1+PV)T_{UU}(P,V)\,,
\end{equation}
where we have dropped the homogeneous part $\phi_{hom}(U,V)$ as before.

\section{ Some formulae}\label{appb}

In the AdS$_2$ space, the bulk-to-boundary propagator 
$K_\Delta$ for conformal dimension $\Delta$ is given by~\cite{Spradlin:1999bn}
\begin{equation} \label{BtoB}
K_{\Delta}(t, r|t') = \langle \chi_R(t,r) \mathcal{O}_R(t') \rangle = 
\frac{2^{\Delta-2}\Gamma^{2}(\Delta)}{\pi\Gamma(2\Delta)}
\Big(\frac{2\pi}{\beta}\Big)^{\Delta}\Big[\frac{r}{L}
- \sqrt{\Big(\frac{r}{L}\Big)^{2}-1}\,  \cosh \frac{2\pi}{\beta}(t-t') \Big]^{-\Delta}\,,
\end{equation}
when $(r,t)$ is spacelike separated from $t'$ at the boundary.
For timelike separation, $t-t'$ should be changed by $t-t'-i\epsilon$ in this
expression. 

It is also useful to rewrite $K_\Delta$ in the global coornates. First we
introduce a distance function
\begin{align}
\sigma(r,t|r',t')
  &= \frac{r}L \frac{r'}L - \sqrt{\Big(\frac{r}{L}\Big)^2-1}  
              \sqrt{\Big(\frac{r'}{L}\Big)^2-1} 
	      \cosh \frac{2\pi}{\beta}(t-t') \notag \\
  &= \frac{\cos(\tau-\tau') - \sin\mu \sin\mu'}{\cos\mu \cos\mu'}\,.
\end{align}
The expression appearing inside the square bracket of \eqref{BtoB} can 
be obtained by taking $r'$ to the right boundary
\begin{align}
\lim_{r'\rightarrow\infty} \frac{L}{r'}\sigma(r,t|r',t') 
&= \lim_{\mu'\rightarrow\pi/2}  \frac{\cos\mu'}{\cos\tau'}\sigma(r,t|r',t')
         \notag \\
&= \frac{\cos(\tau-\tau') - \sin\mu}{\cos\mu \cos\tau'}\,,
\end{align}
giving
\begin{equation} 
K_{\Delta}(t, r|t')=
\frac{2^{\Delta-2}\Gamma^{2}(\Delta)}{\pi\Gamma(2\Delta)}
\Big(\frac{2\pi}{\beta}\Big)^{\Delta}\left[
\frac{\cos(\tau-\tau') - \sin\mu}{\cos\mu \cos\tau'}
\right]^{-\Delta}\,.
\end{equation}

In order to calculate the propagator 
$ \langle \chi_R(t,r) \mathcal{O}_L(-t') \rangle $ with the KMS condition
\eqref{kms}, we need to consider
\begin{equation}
\lim_{r'\rightarrow\infty} \frac{L}{r'} \sigma(r,t|r',-t'+i\beta/2)
  = \frac{\cos(\tau-\tau') + \sin\mu}{\cos\mu \cos\tau'}\,.
\end{equation}
Then
\begin{equation} 
K_{\Delta}(t, r|-t+i\beta/2)=
\frac{2^{\Delta-2}\Gamma^{2}(\Delta)}{\pi\Gamma(2\Delta)}
\Big(\frac{2\pi}{\beta}\Big)^{\Delta}\left[
\frac{\cos(\tau-\tau') + \sin\mu}{\cos\mu \cos\tau'}
\right]^{-\Delta}\,.
\end{equation}
For timelike separation, 
$\langle [ \mathcal{O}_R(t_1), \chi_R(t,r) ] \rangle $ in \eqref{ftt}
becomes
\begin{align}
\langle [ \mathcal{O}_R(t_1), \chi_R(t,r) ] \rangle 
&=K_\Delta (t+i\epsilon,r|t_1) - K_\Delta (t-i\epsilon,r|t_1) \notag \\
&=2i \sin\pi\Delta K_\Delta^r (t-t_1)\,.
\end{align}
where $K^{r}_{\Delta}$ is the retarded function
\begin{align} \label{retarded}
	K^{r}_{\Delta}(t,r|t') &= 
|K_{\Delta}(t,r)|\theta(t) 
\theta\left(\sqrt{\left(\frac{r}{L}\right)^2-1}\cosh\frac{2\pi t}\beta
		-\frac{r}{L} \right) \notag \\
&= \frac{2^{\Delta-2}\Gamma^{2}(\Delta)}{\pi\Gamma(2\Delta)}
\Big(\frac{2\pi}{\beta}\Big)^{\Delta}\left[
\frac{\sin\mu - \cos(\tau-\tau')}{\cos\mu \cos\tau'}
\right]^{-\Delta}
\theta(\tau-\tau')
\theta\Big(\sin\mu-\cos(\tau-\tau')\Big)\,.
\end{align}

By using the formula of the one-loop expectation value of stress tensor~\cite{Birrell:1982ix} from the 2-point function  $G(x,x')$ 
\begin{equation} \label{}
 T_{ab}   = \lim_{x'\rightarrow x}\Big[ \partial_{a}\partial'_{b}G (x,x') -\frac{1}{2}g_{ab}g^{\rho\sigma}\partial_{\rho}\partial'_{\sigma}G(x,x') - \frac{1}{2}g_{ab}M^{2}G(x,x')\Big]\,,
\end{equation}
it is tedious but straightforward to obtain $T_{UU}$  for the expression (\ref{phiTexp}).
Explicitly, the steps go as follows. Firstly, using the following trick~\cite{Gao:2016bin}
\begin{equation} \label{}
\lim_{U',V'\rightarrow U,V}\partial_{U'}\partial_{U} F(U, U') =   \partial_{U}\Big\{\lim_{U' \rightarrow U} \partial_{U'} F(U,U')\Big\}  -  \lim_{U' \rightarrow U} \partial^{2}_{U'} F(U,U')\,,
\end{equation}
one may set 
\begin{equation} \label{}
T_{UU}(P,V) = \partial_{P}H_{1}(P) -  H_{2}(P)\,,
\end{equation}
where $H_{1,2}(U)$ are defined by
\begin{equation} \label{}
H_{1}(U) \equiv 2\lim_{U' \rightarrow U} \partial_{U'} F(U,U')\,, \qquad  
H_{2}(U)  \equiv 2\lim_{U'\rightarrow U}\partial^{2}_{U'}F(U,U')\,.
\end{equation}
A straightforward computation from equation~(\ref{FUU}) leads to
\begin{align}  
H_{1}(U) &=-\int^{U}_{U_{0}}d{\cal S}~ {\cal G}(U,{\cal S})\,,  \label{Hone} \\
H_{2}(U) &=\int^{U}_{U_{0}}d{\cal S}~ {\cal G}(U,{\cal S})\frac{1}{1+UV}\Big[(\Delta+1)\frac{{\cal S}-V}{U{\cal S}+1} + 2V\Big]\,,  \label{Htwo}
\end{align}
where ${\cal G}$ is defined by
\begin{align}   \label{}
{\cal G}(U,{\cal S})   &\equiv  2h\Delta N_{\Delta}   \frac{{\cal S}^{\Delta-1} (1+UV)^{2\Delta-1}({\cal S}-V)^{1-\Delta}}{(U{\cal S}+ 1 )^{\Delta+1}(1+{\cal S}V)^{\Delta} (U/{\cal S}-1)^{\Delta}}\,.
\end{align}

Now, one may note that the dilaton field in (\ref{phiTexp}) becomes
\begin{align}   \label{}
\varphi (U,V) & = \frac{1}{1+UV}\int^{U}_{U_{0}}dP (1+PV)(U-P)\Big[-\partial_{P}H_{1}(P)+H_{2}(P)\Big]  \nn
&= \frac{1}{1+UV}\int^{U}_{U_{0}}dP\Big[(1+PV)(U-P)H_{2}(P) + \Big\{ V(U-P) - (1+PV) \Big\}H_{1}(P)\Big]\,, \nonumber
\end{align}
where we used $H_{1}(U_{0})=0$. Inserting the expressions of $H_{1,2}(P)$ in~(\ref{Hone}) and~(\ref{Htwo}) to the above expression of the dilaton field, one can show that 
\begin{equation} \label{varphi12}
\varphi(U,V)  =  \frac{1}{1+UV}\, \phi_{1}(U,V) + \phi_{2} (U,V)\,, 
\end{equation}
where
\begin{align}   \label{}
\phi_{1}&= (\Delta+1) \int^{U}_{U_{0}}dP\int^{P}_{U_{0}}d{\cal S}~   {\cal G}(P,{\cal S}) \frac{(U-P)({\cal S}-V)}{P{\cal S}+1}  \,, \nn
\phi_{2} & =\int^{U}_{U_{0}}dP\int^{P}_{U_{0}}d{\cal S}~{\cal G}(P,{\cal S}) \,.
\end{align}
One can further simplify  the above expression by noting that $\phi_{1,2}$ could be organized as
\begin{align}   \label{}
\phi_{1} &=    2h\Delta(\Delta +1)N_{\Delta} \int^{U}_{U_{0}}dP\int^{P}_{U_{0}}d{\cal S}~ \frac{{\cal S}^{\Delta-1}({\cal S}-V)^{2-\Delta}}{(1+{\cal S}V)^{\Delta}} {\cal I}_{1}({\cal S},P)\,,    \nn
\phi_{2} &=  2h\Delta N_{\Delta} \int^{U}_{U_{0}}dP\int^{P}_{U_{0}}d{\cal S}~  \frac{{\cal S}^{\Delta-1}({\cal S}-V)^{1-\Delta}}{(1+{\cal S}V)^{\Delta}}  {\cal I}_{2}({\cal S},P)\,,
\end{align}
where we have defined  
\begin{align}   \label{}
{\cal I}_{1} &\equiv \frac{(U-P)(1+PV)^{2\Delta-1}}{(P{\cal S}+1)^{\Delta+2}(P/{\cal S}-1)^{\Delta}}\,, \nn
{\cal I}_{2} &\equiv \frac{(1+PV)^{2\Delta-1}}{(P{\cal S}+1)^{\Delta +1}(P/{\cal S}-1)^{\Delta}}\,.
\end{align}

Let us compute $\phi_{2}$, firstly.  By changing the integration order as 
\begin{equation} \label{COIo}
\int^{U}_{U_{0}}dP\int^{P}_{U_{0}}d{\cal S}    = \int^{U}_{U_{0}}d{\cal S}\int^{U}_{{\cal S}}dP\,,
\end{equation}
and then by the change of a variable
\begin{equation} \label{COVa}
\eta \equiv \frac{1/P-1/U}{1/{\cal S}-1/U}\,,
\end{equation}
and  one can show that
\begin{align}   \label{}
 \int^{U}_{{\cal S}}dP~  {\cal I}_{2} ({\cal S},P)  
= &~  \textstyle{\big[\frac{1}{{\cal S}}-\frac{1}{U}\big]^{1-\Delta}  \big[\frac{1}{U}+{\cal S}\big]^{-\Delta-1} \big[\frac{1}{U} + V\big]^{2\Delta-1} }   \nn
& \times \textstyle{ \int^{1}_{0}d\eta~ [1-\eta]^{-\Delta}\big[1-\frac{1-U/{\cal S}}{1+UV}\eta\big]^{2\Delta-1}\big[1-\frac{1-U/{\cal S}}{1+U{\cal S}}\eta\big]^{-\Delta-1}  } \nn
=&~ \textstyle{ \big[\frac{1}{{\cal S}}-\frac{1}{U}\big]^{1-\Delta}  \big[\frac{1}{U}+{\cal S}\big]^{-\Delta-1} \big[\frac{1}{U} + V\big]^{2\Delta-1} }  \nn 
&  \times \textstyle{\frac{\Gamma(1-\Delta)}{\Gamma(2-\Delta)} F_{1}(1\,;\, 1-2\Delta,1+\Delta\,;\, 2-\Delta\, |\, \frac{1-U/{\cal S}}{1+UV}, \frac{1-U/{\cal S}}{1+U{\cal S}}) }  \,,
\end{align}
where $F_{1}(\alpha\,;\, \beta,\beta'\,;\, \gamma\,|\,x,y)$ denotes the Appell hypergeometric function (See~\cite{Schlosser:2013hbz} for a brief introduction of Appell functions).  One may note  the relation of  the Appell hypergeometric function to the ordinary hypergeometric function, which holds when its arguments satisfy $\beta+\beta'=\gamma$,  as
\begin{equation} \label{Rel1}
F_{1}(\alpha\,;\, \beta,\beta'\,;\, \beta+\beta'\,|\,x,y) =  \frac{1}{(1-y)^{\alpha}} F\Big(\alpha,\beta\,;\,\beta+\beta'\,\Big|\, \frac{x-y}{1-y} \Big)\,.
\end{equation}
As a result, one can see that
\begin{equation} \label{phitwo}
\phi_{2}  = 2h\Delta N_{\Delta} \int^{U}_{U_{0}}d{\cal S} \frac{{\cal S}^{2\Delta-1}}{(1+{\cal S}^{2})^{2\Delta}}\frac{w^{1-\Delta}}{1-\Delta}F(1-\Delta,1-2\Delta\,;\,2-\Delta\,|\,  w)\,,
\end{equation}
where  $w$ is defined by
\begin{equation} \label{Argw}
w\equiv \frac{({\cal S}-V)(U-{\cal S})}{(1+{\cal S}V)(1+{\cal S}U)}\,.
\end{equation}
Note that $w$ is symmetric under the exchange of $U$ and $V$. 


By the same change of the integration order in (\ref{COIo}) and the change of a variable in (\ref{COVa}), one obtains
\begin{align}   \label{}
\int^{U}_{{\cal S}}dP~  {\cal I}_{1} ({\cal S},P)  
= &  \textstyle{\frac{{\cal S}^{\Delta-2}U^{2}(U-{\cal S})^{2-\Delta}}{(1+UV)^{1-2\Delta}(1+U{\cal S})^{\Delta +2}} } \nn 
& \times \textstyle{ \frac{\Gamma(1-\Delta)}{\Gamma(3-\Delta)}~ F_{1}\Big(2\,;\,1-2\Delta,2+\Delta\,;\,3-\Delta\, \Big|\, \frac{1-U/{\cal S}}{1+UV},\frac{1-U/{\cal S}}{1+U{\cal S}} \Big) } \nonumber \\
=& \textstyle{ \frac{(U-{\cal S})^{2-\Delta}}{(1+UV)^{1-2\Delta}(1+U{\cal S})^{\Delta}} \frac{\Gamma(1-\Delta)}{\Gamma(3-\Delta) } \frac{{\cal S}^{\Delta}}{(1+{\cal S}^{2})^{2}}} F\Big(2,1-2\Delta\,;\,3-\Delta\, \Big|\, -\frac{({\cal S}-V)(U-{\cal S})}{(1+UV)(1+{\cal S}^{2})}\Big)\,, \nonumber
\end{align}
where we have used the relation in (\ref{Rel1}). By using the property of the hypergeometric function
\begin{equation} \label{}
F(a,b\,;\, c \,|\, z) = (1-z)^{-b}F\Big(c-a,b\,;\, c\,\Big|\, \frac{z}{z-1}\Big)\,,
\end{equation}
one can show that
\begin{align}  \label{}
\int^{U}_{{\cal S}}dP~  {\cal I}_{1} ({\cal S},P)  
&=    
\frac{{\cal S}^{\Delta}}{(1+{\cal S}^{2})^{2\Delta+1}} \frac{(U-{\cal S})^{2-\Delta}}{(1+U{\cal S})^{1-\Delta}(1+{\cal S}V)^{1-2\Delta}} \nn
& \times \frac{\Gamma(1-\Delta)}{\Gamma(3-\Delta)}  F(1-\Delta,1-2\Delta\,;\, 3-\Delta\,|\, w) \,,
\end{align} 
where $w$ has been introduced in (\ref{Argw}). Now, one can see that 
\begin{align}  \label{phione}
\phi_{1} &=  2h\Delta(\Delta+1)N_{\Delta}(1+UV)\nn 
&\times \int^{U}_{U_{0}}d{\cal S}\frac{{\cal S}^{2\Delta-1}}{(1+{\cal S}^{2})^{2\Delta}}\frac{\Gamma(1-\Delta)}{\Gamma(3-\Delta)}\frac{w^{2-\Delta}}{1-w}F(1-\Delta,1-2\Delta\,;\,3-\Delta\,|\, w) \,.
\end{align} 

Inserting the resultant expressions of $\phi_{1,2}$ given in~\eqref{phione}
and \eqref{phitwo} into  the expression  of the dilaton field $\varphi$ in~(\ref{varphi12}) and using the following relation among hypergeometric functions given by
\begin{align}   \label{}
& c(c-1)(w-1)F(a,b\,;\, c-1\,|\, w) + c[c-1-(2c-a-b-1)w]F(a,b\,;\,c\,|\, w) \nn
& + (c-a)(c-b)wF(a,b\,;\,c+1\,|\, w) =0\,,
\end{align}
one  obtains, finally, 
\begin{equation} \label{}
\varphi(U,V) = 2h\Delta N_{\Delta} \int^{U}_{U_{0}}d{\cal S}\frac{{\cal S}^{2\Delta-1}}{(1+{\cal S}^{2})^{2\Delta}}\Big[ w^{1-\Delta}(1-w)^{2\Delta-1} + \Delta \frac{1+w}{1-w}B_{w}(1-\Delta,2\Delta)\Big]\,,
\end{equation}
where $w$ is given in (\ref{Argw}) and $B_{w}(a,b)$ denotes the incomplete Beta function~\cite{grad}
$$
B_{w}(a,b) = \int^{w}_{0}dt~ t^{a-1}(1-t)^{b-1} = \frac{w^{a}}{a}F(a,1-b\, ;\, 1+a\, |\, w)\,.
$$ 
We would like to emphasize that the integrand in the above expression  of $\varphi(U,V)$ is symmetric in $U$ and $V$ because $w$ is symmetric in those variables. The asymmetry in $U$ and $V$ comes from the integration, which corresponds to the initial condition in our setup. When the left-right boundary interaction is turned off at $t=t_{1} (t_1 \ge t_0)$, one can take the integration range over $[U_{0},U_{1}]$ instead of $[U_{0},U]$ in the above expression as can be inferred from the fact that the turn-off effect could be incorporated as the subtraction of the same integral expression over the range $[U_{1},U]$ if $U > U_1$. In the following, we focus on this case for definiteness.

Noting the identity of the incomplete Beta function
\begin{equation} \label{}
B_{w}(a,b) - B(a,b) = - B_{1-w}(b,a)\,,
\end{equation}
one may observe that the integrand for the dilaton field expression could be written as 
\begin{align}    \label{}
&  w^{1-\Delta}(1-w)^{2\Delta-1}  + \Delta \frac{1+w}{1-w}B_{w}(1-\Delta,2\Delta)  \nonumber \\ 
& =\Delta B(1-\Delta,2\Delta)\frac{1+w}{1-w} + w^{1-\Delta}(1-w)^{2\Delta-1}  - \Delta \frac{1+w}{1-w}B_{1-w}(2\Delta,1-\Delta)\,.
\end{align}
From the definition of $w$ in (\ref{Argw}), it is also useful to  note that  
\begin{equation} \label{}
\frac{1+w}{1-w} = \frac{U+V}{1+UV}\, \frac{2{\cal S}}{1+{\cal S}^{2}} +  \frac{1-UV}{1+UV}\, \frac{1-{\cal S}^{2}}{1+{\cal S}^{2}}\,.
\end{equation}
Using the above observations, one can see that  the dilaton field  expression is given by
\begin{equation} \label{DilDef}
\varphi =  2h\Delta N_{\Delta}\big[  \varphi_{1}   +  \varphi_{2} + \varphi_{3} \big] \,,
\end{equation}
where $\varphi_{i}$'s are defined by
\begin{align}  
\varphi_{1} & \equiv \Delta B(1-\Delta,2\Delta)\frac{U+V}{1+UV}\int^{U_{f}}_{U_{0}} d{\cal S}~  \frac{2{\cal S}^{2\Delta}}{(1+{\cal S}^{2})^{2\Delta+1}}\,, \nn
\varphi_{2}  & \equiv   \Delta B(1-\Delta,2\Delta)\frac{1-UV}{1+UV}\int^{U_{f}}_{U_{0}} d{\cal S}~  \frac{{\cal S}^{2\Delta-1}(1-{\cal S}^{2})}{(1+{\cal S}^{2})^{2\Delta+1}}\,, \nonumber \\
\varphi_{3} & \equiv  \int^{U_{f}}_{U_{0}} d{\cal S}~  \frac{{\cal S}^{2\Delta-1}}{(1+{\cal S}^{2})^{2\Delta}} \Big[w^{1-\Delta}(1-w)^{2\Delta-1}  - \Delta \frac{1+w}{1-w}B_{1-w}(2\Delta,1-\Delta)\Big]\,.  \label{horCh}
\end{align} 
Here,  $U_f = U_1$ if $U \ge U_1$ and $U_f =U$ if $U_1 >  U \ge U_0$. 
By using the integral representation of the incomplete Beta function
\begin{equation} \label{}
\int^{\infty}_{U}d{\cal S}\frac{{\cal S}^{2\Delta +m-1}}{(1+{\cal S}^{2})^{2\Delta + n -1}} = \frac{1}{2}B_{z}(\Delta-\frac{m}{2}+n -1,\Delta + \frac{m}{2})\,,  \qquad  z \equiv \frac{1}{1+U^{2}}\,,
\end{equation}
one can see that  $\varphi_{1,2}$ are given by the closed forms as follows 
\begin{align}    \label{}
\varphi_{1} & =\Delta B(1-\Delta,2\Delta)\,   \frac{U+V}{1+UV} \Big[ B_{z_{0}}(\Delta+\frac{1}{2},\Delta+\frac{1}{2})- B_{z_{f}}(\Delta+\frac{1}{2},\Delta+\frac{1}{2})\Big]\,,   \nn
\varphi_{2}&= \frac{1}{2} \Delta B(1-\Delta,2\Delta)\, \frac{1-UV}{1+UV}\Big[ \Big\{B_{z_{0}}(1+\Delta,\Delta) - B_{z_{f}}(1+\Delta,\Delta)  \Big\}  \nn
& \qquad \qquad  \qquad \qquad \qquad \qquad - \Big\{  B_{z_{0}}(\Delta,1+\Delta)- B_{z_{f}}(\Delta,1+\Delta) \Big\}\Big]   \nonumber \qquad \\
& = \frac{1}{2}B(1-\Delta,2\Delta)\,  \frac{1-UV}{1+UV} \Big[  z^{\Delta}_{f}(1-z_{f})^{\Delta} - z^{\Delta}_{0}(1-z_{0})^{\Delta}  \Big]\,,  \qquad  
\end{align}
where  in the last equality we have used the incomplete Beta function relation given by
\begin{equation} \label{}
B_{x}(\Delta,1+\Delta) - B_{x}(1+\Delta,\Delta) = \frac{1}{\Delta} x^{\Delta}(1-x)^{\Delta}\,.
\end{equation}

In summary, one obtains 
\begin{equation} \label{}
\varphi(U,V) =  h \alpha_{s}~ \frac{U+V}{1+UV}   + h \alpha_{c}~ \frac{1-UV}{1+UV} + 2h\Delta N_{\Delta}\, \varphi_{3}(U,V)\,,
\end{equation}
where  $\varphi_{3}$ is given in~(\ref{horCh}) and,   $\alpha_{s}$ and $\alpha_{c}$ are defined by
\begin{align} 
\alpha_{s} &=  2\Delta^{2} N_{\Delta}B(1-\Delta,2\Delta)~ \Big[ B_{z_{0}}\Big(\Delta+\frac{1}{2},\Delta+\frac{1}{2}\Big)- B_{z_{f}}\Big(\Delta+\frac{1}{2},\Delta+\frac{1}{2}\Big)\Big]
 \nn
 &= \frac{1}{2\pi}\frac{2^{2\Delta-1}\Delta\Gamma^{2}(\Delta)}{\Gamma(2\Delta)}~ \Big[ B_{z_{0}}\Big(\Delta+\frac{1}{2},\Delta++\frac{1}{2}\Big)- B_{z_{f}}\Big(\Delta+\frac{1}{2},\Delta+\frac{1}{2}\Big)\Big]\,, \label{b40} \\ 
 \alpha_{c} &=  \Delta N_{\Delta}B(1-\Delta,2\Delta)~  \Big[  z^{\Delta}_{f}(1-z_{f})^{\Delta} - z^{\Delta}_{0}(1-z_{0})^{\Delta}  \Big] \nn
&=  \frac{1}{2\pi}\frac{2^{2\Delta-2}\Gamma^{2}(\Delta)}{\Gamma(2\Delta)}~ 
 \left[ \bigg(  \frac{U_{f}}{1+U^{2}_{f}}\bigg)^{2\Delta}
-  \bigg(  \frac{U_{0}}{1+U^{2}_{0}}\bigg)^{2\Delta}  \right]\,. \label{DilExp}
\end{align}
Here, we would like to emphasize that $\alpha_{s}$ is always positive semi-definite, while $\alpha_{c}$ could be negative.

Let us consider the region 
where $UV \simeq -1$, 
 the expansion of $1-w$ is given by 
\begin{equation} \label{}
1-w=  \frac{1+UV}{U}\frac{1+{\cal S}^{2}}{{\cal S}} \bigg[1+ \Big(\frac{1}{{\cal S}U}+{\cal S}V+ \frac{V}{U}\Big)+ \cdots \bigg]\,,
\end{equation}
and so, through the expansion of the incomplete Beta function $B_{x}(a,b)$ around $x=0$,  the integrand of $\varphi_{3}$ could be expanded as
\begin{equation} \label{}
w^{1-\Delta}(1-w)^{2\Delta-1}  - \Delta \frac{1+w}{1-w}B_{1-w}(2\Delta,1-\Delta)  = -\frac{1}{2\Delta(1+2\Delta)} (1-w)^{2\Delta}+ \cdots\,.
\end{equation}
Now, one can notice that the integrand expression for $\varphi_{3}$ becomes very small  in the limit  and so $\varphi_{3}$ term could be ignored.  As a result,  one can write
\begin{equation} \label{}
\varphi(U,V) =  h \alpha_{s}~ \frac{U+V}{1+UV}   + h \alpha_{c}~ \frac{1-UV}{1+UV} +{\cal O}\Big[\, (1+UV)^{2\Delta} \Big]\,.
\label{kru}
\end{equation}
Finally, note that the higher order corrections  can be written as
\begin{equation} \label{}
{\cal O}(\cos^{2\Delta} \mu) 
\end{equation}
by using the fact $(1+UV) \propto \cos \mu$.

\section{ Dilaton Deformation in Kruskal coordinates}
In this appendix we consider the position of the singularity on the left/right wedges in the Kruskal coordinates. 
Recall that the position of the black hole singularity could read from  $\phi(U,V)=0$.  Before the deformation, it is given by $UV=1$, 
as can be seen from  the expressions of the homogeneous solution $\phi_{hom}(U,V)$ in Kruskal coordinates 
\begin{equation} \label{}
\phi_{hom}(U,V)=
L \bigg[\frac{1-UV}{1+UV}\bigg]\,.
\end{equation}
%
By adding the back-reacted deformation effect on the dilaton field, one obtains
\begin{equation} \label{BackDil}
\phi_{hom}(U,V) + \varphi(U,V) = \Big(L + h\alpha_{c}\Big) \bigg[\frac{1-(U-h\alpha_{s}/L)(V-h\alpha_{s}/L)}{1+UV}\bigg] + {\cal O}(h^{2})\,,
\end{equation}
where $\alpha_{s}$ and $\alpha_{c}$ could be read from (\ref{DilExp}).
%
Now, one can see that the position of singularity in the left/right wedges is determined by  $1-(U-\alpha_{c}/L)(V-\alpha_{s}/L) =0$.  If we focus on the position of the singularity on the left/right boundary, we can see that it is given by  $(U\simeq h\alpha_{s}/L,~ V\rightarrow\infty)_{L}$ / $(U\rightarrow \infty,~ V\simeq h\alpha_{s}/L)_{R} $. This shows us that the position of the singularity is moved upward slightly   as far as  $h >0$,  compared to the undeformed cases which are given by $(U=0,~ V\rightarrow\infty)_{L}$ / $(U\rightarrow \infty,~ V=0)_{R} $, respectively.

\section{The case of $b\neq 0$ and $\tau_{B}\neq 0$ }
The relation bewteen the Kruskal coordinates and the global ones in the case of $b\neq 0$ $\tau_{B}\neq 0$ is given by
\begin{align}    \label{}
\frac{1-UV}{1+UV}&= \frac{r}{L} = \frac{b+b^{-1}}{2}\frac{\cos(\tau-\tau_{B})}{\cos\mu} -  \frac{b-b^{-1}}{2} \tan\mu \,, \nn 
\frac{U+V}{1+UV}&= \sqrt{\frac{r^{2}}{L^{2}}-1}~ \sinh\frac{L}{\ell^{2}}t =  \frac{\sin(\tau - \tau_{B})}{\cos\mu}\,,
\end{align}
which leads to the following form of the  backreaction to the dilaton field
\begin{equation} \label{}
\varphi(\tau,\mu) =  \frac{1}{\cos\mu} \Big[h \alpha_{s} \sin (\tau -\tau_{B})  + \frac{b+b^{-1}}{2}\, h \alpha_{c}\cos (\tau-\tau_{B}) -   \frac{b- b^{-1}}{2}\, h\alpha_{c}\sin\mu \Big] + \cdots\,.
\end{equation}
%

\section{Relation to the boundary action}\label{eee}
\renewcommand{\theequation}{E.\arabic{equation}}
  \setcounter{equation}{0}
In this section, we will set $\ell=L=8\pi G=1$.
The  effective boundary action corresponding the bulk action is known to be given by~\cite{Maldacena:2017axo,Maldacena:2018lmt}
\begin{equation} \label{}
S = \int d\tilde{u} \bigg[ -\phi_{l}\Big\{ \tan \frac{\tau_{l}(\tilde{u})}{2},\tilde{u}\Big\}  -\phi_{r}\Big\{ \tan \frac{\tau_{r}(\tilde{u})}{2},\tilde{u} \Big\}  + \frac{g}{2^{2\Delta}}\bigg(\frac{\tau'_{l}(\tilde{u})\tau'_{r}(\tilde{u})}{\cos^{2}\frac{\tau_{l}(\tilde{u})-\tau_{r}(\tilde{u})}{2}} \bigg)\bigg]\,,
\end{equation}
where  $\phi_{l}=\phi_{r}$ can be identified with $\bar{\phi}$ in the bulk.  In the following, we would like to clarify the relation of  the boundary time $\tilde{u}$ and the bulk Rindler wedge  time $t$ in  Eq.~(\ref{btz}).  More correctly, there are left/right Rindler wedge times $t_{l/r}$, while $\tilde{u}$ denotes the simultaneous intrinsic  boundary time in both boundaries. In our setup, one may set   $t(\tilde{u}) \equiv t_{r}(\tilde{u}) = -t_{l}(\tilde{u})+{\cal O}(g)$. By using the relation between the global time $\tau$ and $t$, one can see that the equations of motion is given, up to the relevant order, by
\begin{equation} \label{SchEOM}
\bar{\phi}\bigg[\frac{1}{t'}\Big(\frac{t''}{t'}\Big)' - t' \bigg]' - \frac{g\Delta(2\Delta-1)}{2^{2\Delta}}\bigg[\frac{t'^{2\Delta-2}t'' - t'^{2\Delta}\tanh t}{\cosh^{2\Delta}t } \bigg] =0 \,,
\end{equation}
where ${}^{'} \equiv \frac{d}{d\tilde{u}}$.
As in the bulk, the coupling $g$ is chosen such that $g= g_{0}\Big[\theta(t-t_{0})- \theta(t-t_{1})\Big]$, and then the coupling $g_{0}$ could be identified with the bulk parameter $h$ with an appropriate numerical factor. For $t < t_{0}$, the solution of the above equations of motion is given by $t(\tilde{u}) = \tilde{u}$. Since the coupling is taken as  $g_{0} \ll 1$, it would be sufficient to consider the perturbative solution,  for  the range $ t_{0} < t < t_{1}$, as
\begin{equation} \label{}
t'(\tilde{u}) =  1 + g_{0} F(\tilde{u}) + {\cal O}(g^{2}_{0})\,.
\end{equation}
Inserting this ansatz to Eq.~(\ref{SchEOM}), one obtains 
\begin{equation} \label{}
\Big[\bar{\phi}(F'' - F) - \frac{2\Delta -1}{2^{2\Delta +1}}\frac{1}{\cosh^{2\Delta}t}\Big]' = 0\,,
\end{equation}
where one may exchange $t(\tilde{u})$ with $\tilde{u}$ since their difference resides in higher orders in $g$. Now,  one can show that the solution of $F$ is given by 
\begin{equation} \label{bdRes}
\bar{\phi} F(\tilde{u}) =  A  \sinh t  +  B \,,
\end{equation}
where
\begin{align}    \label{}
A &\equiv  \Delta \Big[B_{z_{0}}\Big(\Delta + \frac{1}{2},\Delta + \frac{1}{2}\Big)- B_{z}\Big(\Delta + \frac{1}{2},\Delta + \frac{1}{2}\Big)  \Big]\,,   \qquad z = \frac{1}{1+U^{2}}= \frac{1}{1+e^{2t(\tilde{u})}}\,, \nonumber \\
B &= \frac{1}{2^{2\Delta+1}} \bigg[ \frac{1}{\cosh^{2\Delta}t}  -  \frac{1}{\cosh^{2\Delta}t_{0}}\bigg]\,.  \nonumber 
\end{align}
The constants in $A$ and $B$ are chosen in such a way that $t'(\tilde{u})$ becomes $t'(\tilde{u})=1$ at $t=t_{0}$.

By recalling the cut-off at the right boundary taken as 
\begin{align}    \label{}
ds^{2}\Big|_{bd} &= -\frac{1}{\epsilon^{2}}d\tilde{u}^{2} = - r^{2}t'^{2}(\tilde{u}) + {\cal O}(1)\,, \nonumber \\
 \phi \Big|_{bd} &= \frac{\bar{\phi}}{\epsilon}=r\Big[\bar{\phi} + h \alpha_{c}+ h\alpha_{s}\sinh t \Big]+{\cal O}\Big(\frac{1}{r^{2\Delta}}\Big)\,,   \nonumber  
\end{align}
one obtains the bulk time $t$ in terms of the boundary time $\tilde{u}$ 
\begin{equation} \label{bulkRes}
t'(\tilde{u}) = 1 + \frac{1}{\bar{\phi}}\Big[ h\alpha_{c} + h\alpha_{s}\sinh t \Big]+ {\cal O}(\epsilon)\,.
\end{equation}
By choosing 
\begin{equation} \label{}
g_{0} = 2 h \Delta N_{\Delta}B(1-\Delta, 2\Delta) = \frac{h}{2\pi}\frac{2^{2\Delta-1}\Gamma^{2}(\Delta)}{\Gamma(2\Delta)}\,,
\end{equation}
one can check that the expressions in Eq.~(\ref{bdRes}) and Eq.~(\ref{bulkRes})  from the boundary and the bulk, respectively,  match completely.

\end{document}